\documentclass[preprintnumbers,showpacs,amsmath,amssymb,floatfix,prd,onecolumn,superscriptaddress,nofootinbib]{revtex4-2}
\usepackage{subfigure}
\usepackage{graphicx}
\usepackage{epsfig}
\usepackage{bm}
\usepackage{amsfonts}
\usepackage{epstopdf}
\usepackage[colorlinks,linkcolor=blue,anchorcolor=blue,citecolor=green]{hyperref}
\usepackage{cancel}
\def\equationautorefname~#1\null{Eq.~(#1)\null}
\newcommand*{\dif}{\mathop{}\!\mathrm{d}}

\begin{document}
	
\title{Image of Kerr-de Sitter black holes illuminated by equatorial thin accretion disks }

\author{Ke Wang}
\affiliation{Division of Mathematical and Theoretical Physics, Shanghai Normal University, 100 Guilin Road, Shanghai 200234,  P.R.China}

\author{Chao-Jun Feng}
\thanks{Corresponding author}
\email{fengcj@shnu.edu.cn}
\affiliation{Division of Mathematical and Theoretical Physics, Shanghai Normal University, 100 Guilin Road, Shanghai 200234,  P.R.China}

\author{Towe Wang}
\email{twang@phy.ecnu.edu.cn}
\affiliation{Department of Physics, East China Normal University, Shanghai 200241, China\\}

\begin{abstract}

To explore the influence of the cosmological constant on black hole images, we have developed a comprehensive analytical method for simulating images of Kerr-de Sitter black holes illuminated by equatorial thin accretion disks. Through the application of explicit equations, we simulate images of Kerr-de Sitter black holes illuminated by both prograde and retrograde accretion disks, examining the impact of the cosmological constant on their characteristic curves, relative sizes, and observed intensities. Our findings reveal that, in comparison to Kerr black holes, the cosmological constant not only diminishes the relative size of a black hole but also amplifies its luminosity. Moreover, an observer's relative position in the universe (\(r_0/r_C\)) can influence both the relative size and luminosity of a black hole, where \(r_0\) is the distance from the observer to the black hole, \(r_C\) is the cosmological horizon determined by the value of the cosmological constant \(\Lambda\).
\end{abstract}


\maketitle

\section{Introduction}

In April 2019, the Event Horizon Telescope (EHT) collaboration released images of the core of the galaxy M87 (M87*) based on 1.3mm interferometric observations 
\cite{EventHorizonTelescope:2019dse,EventHorizonTelescope:2019uob,EventHorizonTelescope:2019jan,
EventHorizonTelescope:2019ths,EventHorizonTelescope:2019pgp,EventHorizonTelescope:2019ggy}.
These images all acknowledge a ring-like structure of typical diameter \(\sim 40\mu \)as. This is important evidence of the existence of black holes, and it raises the intriguing prospect of testing general relativity with future high-resolution black hole images.

The basic feature of black hole images is a bright ring surrounding a dark area \cite{lu_ring-like_2023}. Early theoretical studies mainly focused on the central dark area of black hole images \cite{synge_escape_1966,bardeen_timelike_1973},
commonly regarded as the `black hole shadow'. Specifically, in numerous previous studies of black hole shadows 
\cite{perlick_calculating_2022,cunha_shadows_2016,narayan_shadow_2019,perlick_black_2018,wang_shadows_2019,huang_revisiting_2018,abdujabbarov_shadow_2016,
Li:2013jra,Abdujabbarov:2012bn,Yumoto:2012kz,Dexter:2012fh,Amarilla:2010zq,Bambi:2010hf,Huang:2007us,Falcke:1999pj,
Cunha:2017eoe,Abdujabbarov:2016hnw,Zhang:2019glo,Guo:2020zmf,Hu:2020usx,Hou:2021okc,Wang:2021ara,Gan:2021xdl,Meng:2022kjs,Yang:2022btw},
the black hole shadow corresponds to the geometric shape of the projection of the photon sphere/shell on an observer's sky, where the photon sphere/shell is a region composed of bound photon orbits. 
Recently there has been some controversy over the term `black hole shadow' \cite{vincent_images_2022,bronzwaer_nature_2021}.
To avoid inaccuracy of terminology, we use the term `critical curve' introduced by \cite{gralla_lensing_2020} to represent the edge of the black hole shadow.
The size and shape of the critical curve can usually be analytically calculated, even for black holes of the Pleb\'{a}nski-Demi\'{a}nski class
(Kerr-de Sitter black holes are a special case of this class.)\cite{plebanski_rotating_1976,griffiths_new_2006,grenzebach_photon_2014,grenzebach_photon_2015}.
Unlike the critical curve, studying the black hole image mostly requires numerical simulation.
The first simulated black hole image presented in \cite{luminet_image_1979} displays a Schwarzschild black hole surrounded by a rotating luminous accretion disk.
Similar equatorial accretion disk models are now widely used to simulate black hole images and are capable of producing similar results to general-relativistic magnetohydrodynamic (GRMHD) simulations \cite{johnson_universal_2020,chael_observing_2021,gold_verification_2020}.

In \cite{Medeiros:2023pns}, the authors employ a cutting-edge dictionary-learning-based algorithm named PRIMO, utilizing high-fidelity simulations of accreting black holes as a training set. By discerning correlations among various regions of the interferometric data space, this methodology facilitates the recovery of high-fidelity images, even in scenarios of sparse coverage, enabling attainment of the nominal resolution of the EHT array.In \cite{Zhang:2023okw}, the authors present that the predictive values for massive black holes exhibit universality within a broad class of quantum-modified spacetime, incorporating the scenario of a black hole transitioning to a white hole. Consequently, their findings introduce a new experimental avenue to scrutinize predictions of quantum gravity. In \cite{Qin:2023nog}, the authors capture the polarized image of a rotating black hole enveloped by a cold dark matter halo. In \cite{Wang:2022mjo}, the author investigates primary images (PIs) and secondary images (SIs) resulting from strong gravitational lensing around a Kerr black hole shadow. In \cite{Pantig:2022qak}, symmergent gravity is evaluated through shadow imaging and weak field photon deflection by a rotating black hole, utilizing results from M87$^*$ and Sgr. $\hbox {A}^*$. In \cite{Shaikh:2022ivr}, the authors test black hole mimickers using the Event Horizon Telescope image of Sagittarius A*. In \cite{Qin:2022kaf}, polarized images of a synchrotron-emitting ring in the spacetime of a rotating black hole within the scalar–tensor–vector–gravity (STVG) theory are explored. In \cite{Hou:2022eev}, the observational appearance of a Kerr-Melvin black hole (KMBH) illuminated by an accretion disk is studied. In \cite{Gan:2021pwu}, the authors investigate null geodesics of a class of charged, spherical, and asymptotically flat hairy black holes in an Einstein-Maxwell-scalar theory with nonminimal coupling for the scalar and electromagnetic fields. In \cite{PhysRevD.107.104041}, the Hubble law is extracted through frequency-shift considerations of test particles revolving around the Kerr black hole in asymptotically de Sitter spacetime. In \cite{Hioki:2009na}, the authors delve into the correspondence between the shape of the shadow and Kerr parameters, extending it to a general rotating black hole generated by the Newman-Janis algorithm, as detailed in Refs. \cite{Tsukamoto:2014tja} and \cite{Tsukamoto:2017fxq}. The authors investigate images of a reduced Kiselev black hole in \cite{Qu:2023rsv} and images of nonsingular nonrotating black holes in conformal gravity in \cite{Qu:2023hsy}. For additional works, refer to the reviews in \cite{Wang:2022kvg} \cite{Perlick:2021aok} and also \cite{Chen:2022scf}.

In principle, a camera aimed at a black hole sees an infinite sequence of self-similar ring-like structures near the critical curve,
the ring-like structure in the black hole image is named `the photon ring' \cite{bardeen_timelike_1973,gralla_black_2019,johnson_universal_2020}.
in which subrings arise from photons that differ by the number of half-orbits they complete around the black hole on the way from their source to the detector.
These subrings may produce strong and universal signatures on long interferometric baselines \cite{johnson_universal_2020}.

The cosmological constant \(\Lambda\), as we know, is the simplest model to explain the cosmic accelerating  expansion, and the corresponding cosmological model is commonly called the $\Lambda$ cold dark matter ( \(\Lambda\)CDM ) model \cite{SupernovaSearchTeam:1998fmf}.
In a universe with a non-zero cosmological constant \(\Lambda\), a rotating black holes could be described by the Kerr-de Sitter (KdS) metric discovered by Carter \cite{carter_black_1973},
and today we know that it is a special case of the general Pleb\'{a}nski-Demi\'{a}nski family of metrics \cite{plebanski_rotating_1976}.

Although the current observations of M87* have limited resolution, and it is anticipated that the impact of the cosmological constant on the black hole image will be minimal,
it remains crucial to elucidate how the cosmological constant may influence the black hole image.
Previous investigations into Kerr-de Sitter black hole shadows have demonstrated that the cosmological constant reduces the size of the critical curve \cite{li_shadow_2020,omwoyo_remarks_2022}.
In this paper, we try to investigate the effects of the cosmological constant on the black hole images and the possibility of using black hole images to test the cosmological constant.
We build upon and extend a simple geometric model proposed by \cite{gralla_shape_2020} in Kerr black hole spacetimes,
which was later simplified by \cite{chael_observing_2021} and employed to fit the GRMHD results.
In this model, the black hole is surrounded by a geometrically and optically thin equatorial accretion disk.
Particles in the accretion disk move along time-like geodesics, while the emission profiles remain stationary and axisymmetric.
We gradually establish imaging methods suitable for this type of model, including analytical ray tracing methods, and finally make the formula for generating black hole images completely explicit.

This paper is organized as follows: In \autoref{2}, we review the Kerr-de Sitter spacetime and establish the equatorial disk model.
In \autoref{3}, we describe the complete method of generating images, in which the analytical ray tracing is discussed in \autoref{4}.
Our results from various simulations are presented in \autoref{5}. Finally, we summarize our work in \autoref{6}.

Throughout the paper we will use geometric units \(c=G=1\) and the convention of metric is \((-,+,+,+)\).

\section{Spacetime of kerr-de Sitter black holes}
\label{2}
In this section, we briefly review the geodesic equations of the Kerr-de Sitter spacetime and the equatorial circular orbits.  After that, an analytic accretion disk model will be described. For a more detailed treatment of equatorial circular orbits, refer to \cite{stuchlik_equatorial_2004}.
\subsection{The spacetime and geodesics}
\label{2.1}

In standard Boyer-Lindquist coordinates \((t,r,\theta,\phi)\), the metric  of Kerr-de Sitter spacetime \cite{carter_black_1973} is 
\begin{equation}
\begin{split}
        \dif s^2=&\frac{a^2\sin ^2\theta \Delta _{\theta }-\Delta _r}{\Xi ^2\Sigma } \dif t^2
        -2\frac{a \sin ^2\theta \left(\Delta _{\theta }\left(a^2+r^2\right)-\Delta_r\right)}{\Xi ^2\Sigma } \dif t \dif \phi
        \\&+\frac{\Sigma }{\Delta _r}\dif r^2+\frac{\Sigma }{\Delta _{\theta }} \dif \theta^2
        +\frac{\sin ^2\theta \left(\Delta _{\theta}\left(a^2+r^2\right){}^2-a^2\Delta _r\sin ^2\theta \right)}{\Xi ^2\Sigma } \dif \phi^2 \,,
\end{split}
\end{equation}
where $M, a$ are the mass and spin of the black hole, $\Lambda >0$ is the cosmological constant, and  
\begin{equation}
    \begin{aligned}
        \Sigma &=r^2+a^2\cos ^2\theta  \,,\\
        \Delta _r&=\left(r^2+a^2\right)\left(1-\frac{\Lambda }{3}r^2\right)-2M r \,,\\
        \Delta _{\theta }&=1+\frac{\Lambda }{3}a^2\cos ^2\theta \,,\\
        \Xi &=1+\frac{\Lambda }{3}a^2 \,.
    \end{aligned}
\end{equation}
In the following, we define a the dimensionless cosmological constant 
\begin{align}
    y =\frac{1 }{3}\Lambda M^2 \,,
\end{align}
then we set \(M=1\), such that the coordinates \(x^\mu\), the spin \(a\) and the line element \(\dif s\) becomes dimensionless.

According to the symmetry of the spacetime, three geodesic constants can be defined: the total energy \(E\), the axial angular momentum \(L\), and the Carter constant \(Q\) \cite{carter_black_1973}. 
\(E\) and \(L\) correspond to the Killing vectors \(\partial_t\) and \(\partial_\phi\), respectively, and \(Q\) is related to the `hidden symmetry' of spacetime.
These three constants, together with the conserved Hamiltonian \(H=\mu/2\), simplify the geodesic equation into four first-order differential equations, which can be compactly written as
\begin{equation}
    \label{eq.geodesic}
    g_{\mu \nu } \dot{x}^{\nu }=p_{\mu }=
    \left(-E,\pm _r\frac{\sqrt{\bar{R}(r)}}{\Delta _r},\pm _{\theta }\frac{\sqrt{  \bar{\Theta }(\theta)}}{\Delta _{\theta }},L\right) \,,
\end{equation}
where  $p_\mu$ is the canonical momentum, \(\pm _r=\text{sgn}(p^r)\), \(\pm _\theta=\text{sign}(p^\theta)\), and we have defined
\begin{equation}
\begin{aligned}
        \bar{\Theta }(\theta )&=-\Xi ^2(L \csc\theta-a E \sin\theta)^2+\left(Q+(a E -L)^2 \Xi ^2+\mu \left(a^2-a^2 \sin^2\theta \right)\right)\Delta _{\theta }  \,,\\
        \bar{R}(r)&=\left(a^2 E-a L+E r^2\right)^2 \Xi ^2-\left(Q+(a E-L)^2 \Xi ^2-r^2 \mu \right) \Delta _r \,.
    \end{aligned} 
\end{equation}
The null geodesics, $\mu =0$, is determined only by two parameters  \(\lambda=L/E\) and \(\eta=Q/E^2\) called the impact parameters. So the null geodesics can be recast in the form of
\begin{equation}
    \begin{aligned}
        \label{nullgeos}
        g_{\mu \nu } \dot{x}^{\nu }=p_{\mu }=
        E\left(-1,\pm _r\frac{\sqrt{R(r)}}{\Delta _r},\pm _{\theta }\frac{\sqrt{ \Theta(\theta ) }}{\Delta _{\theta }},\lambda \right) \,,
    \end{aligned}
\end{equation}
where
\begin{equation}
    \begin{aligned}
        \label{RRandTheta}
        \Theta (\theta )&=\Delta _{\theta } \left(\Xi ^2 (a-\lambda )^2+\eta \right)-\Xi ^2 (\lambda  \csc \theta -a \sin \theta )^2 \,,
        \\R(r)&=\Xi ^2\left(a^2-a \lambda +r^2\right)^2-\Delta _r\left(\Xi ^2(\lambda -a)^2+\eta \right)\,.
    \end{aligned}
\end{equation}

\subsection{Equatorial timelike geodesics}
\label{2.2}
The timelike geodesics on equatorial plane possess conserved quantities \(Q=0\) and \(\mu=-1\), then the radius function reduced from \autoref{eq.geodesic} is
\begin{equation}
    \Sigma ^2\dot{r}^2=\left(a^2 E-a L+E r^2\right)^2 \Xi ^2-\left((L-a E)^2 \Xi ^2+r^2 \right) \Delta _r\equiv V(r) \,.
\end{equation}
The equatorial circular orbits require \(V(r)=V'(r)=0\),
 whose solutions obtained by \cite{stuchlik_equatorial_2004} 
are
\begin{equation}
    \label{EandLofE}
        E_{\pm } \Xi =\frac{1-\frac{2}{r}-\left(r^2+a^2\right)y \pm a\left(\frac{1}{r^3}-y\right)^{1/2}}{\sqrt{1-\frac{3}{r}-a^2y\pm 2a\left(\frac{1}{r^3}-y\right)^{1/2}}} \,,
\end{equation}
\begin{equation}
    \label{EandLofL}
        L_{\pm } \Xi =\frac{-2a-a r\left(r^2+a^2\right)y\pm r\left(a^2+r^2\right)\left(\frac{1}{r^3}-y\right)^{1/2}}{r\sqrt{1-\frac{3}{r}-a^2y\pm 2a\left(\frac{1}{r^3}-y\right)^{1/2}}} \,,
\end{equation}
where $\pm$ denotes  the co-rotating orbit \((aL>0)\) and counter-rotating orbits \((aL<0)\), respectively.

For \(0< y \ll1\), the stability of circular orbits is straightforward to analyze.
The stability requirement \(V''(r) \geq 0\) determines the range of the radii of the stable circular orbits, which is equivalent to
\begin{align}
    \label{eq:v''r0}
    \frac{8 a r^2 \sqrt{\frac{1}{r^3}-y} \left(r^3 y-1\right)\pm r \left(6-r-15 r^3 y+4 r^4 y\right) \pm a^2 \left(3+r^2 y-4 r^5 y^2\right)}
    {2 a r \sqrt{\frac{1}{r^3}-y}\pm \left(r-3-a^2 r y\right)}\leq 0 \,.
\end{align}
Outside the horizon, this condition can be reduced to
\begin{align}
    r_\text{ISCO}<r<r_\text{OSCO} \,,
\end{align}
which defines the radii of the innermost stable circular orbit (ISCO) and the outermost stable circular orbit (OSCO).  It is difficult to obtain the analytic form of  \(r_\text{ISCO}\) and \(r_\text{OSCO}\) since \autoref{eq:v''r0} is related to high-order polynomials of \(r\).

\begin{figure}[htbp]
    \centering
    \includegraphics[width=0.495\textwidth]{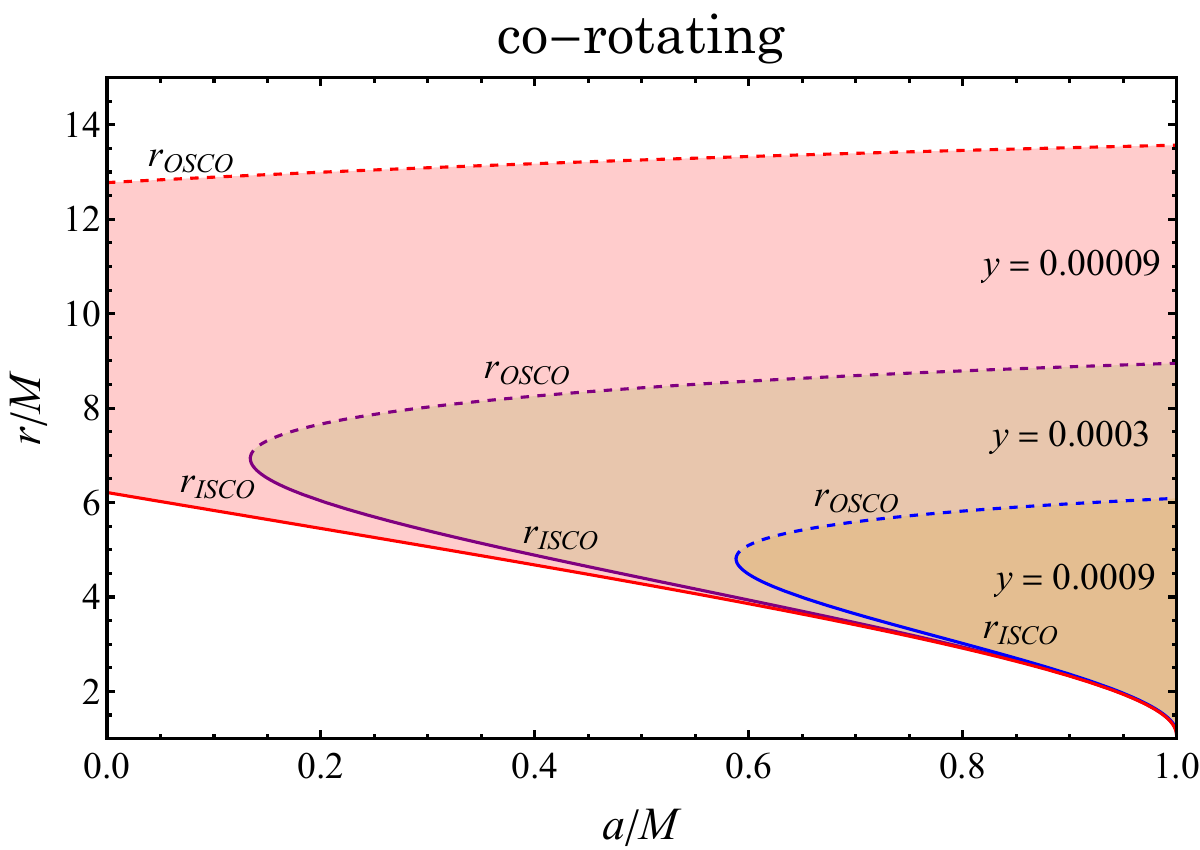}
    \includegraphics[width=0.495\textwidth]{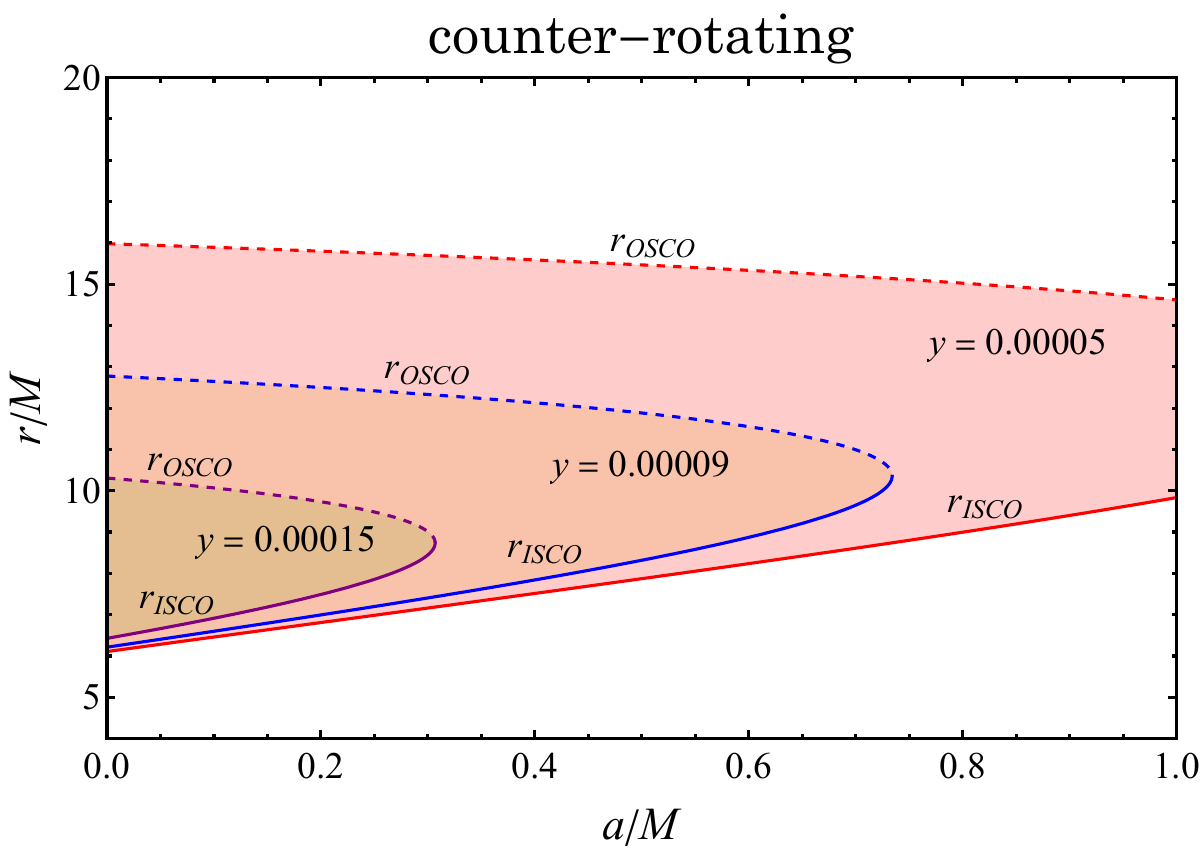}

    \caption{The allowable region of the stable equatorial circular orbits in parameter \(r\wedge a\) space. 
    The solid curves and dashed curves represent \(r_\text{ISCO}(a)\) and \(r_\text{OSCO}(a)\) at different values of \(y\), respectively.
    (Left) In the case of co-rotating orbits, from the outside to the inside, 
    the solid or dashed curves with color (red, purple, blue) correspond to the values of \(y\) at (0.00009, 0.0003, 0.0009).
    (Right) In the case of counter-rotating orbits, from the outside to the inside,
    the solid or dashed curves with color (red, blue, purple) correspond to the values of \(y\) at (0.00005, 0.00009, 0.00015). }

    \label{fig:co-counter-rotating}
\end{figure}

In the limit of \(y\to0\), \autoref{eq:v''r0} reduces to the condition of stable circular orbits in the Kerr spacetime.  In such case, the OSCO does not exist, in other words, \(\lim_{y \to 0}  r_{\text{OSCO}}= \infty\).  For a succinct analytical solution for \(r_\text{ISCO}\) in the Kerr spacetime, see \cite{bardeen_rotating_1972}.
When $y\neq 0$, the range of stable orbits will shrink as $y$ increases for a fixed value of $a$, which  is reflected in an increase of \(r_\text{ISCO}\) and a decrease of \(r_\text{OSCO}\).
And eventually, \(r_\text{ISCO}\) and \(r_\text{OSCO}\) will be equal and then cross at a specific value of \(y\), at which point the stable orbits disappear. However, when $y$ is fixed, the stability range for co-rotating orbits will expand as $a$ increases, while that for counter-rotating will shrink. At a specific value of \(a\),  the stability range can be also vanished. In  \autoref{fig:co-counter-rotating}, it shows how the allowable region of the stable equatorial circular orbits varies with \(a\) and \(y\). In the following, we will assume that \(y\ll1\) 
\footnote{In Kerr-de Sitter spacetimes, the cosmological horizon \(r_C\) is governed by the cosmological constant as \(r_C \approx  \sqrt{1/y}-1\). Since the observer is within the cosmological horizon, the cosmological horizon must be greater than the observation distance to conform to current assumptions about the size of the universe, which requires the cosmological constant to be much less than 1.}
, such that the ISCO and OSCO are separated far enough to allow for a stable equatorial accretion disk outside the black hole. 

\subsection[short]{The thin disk model}
\label{2.3}

The phenomenological model for M87* established in Kerr spacetime is proposed in \cite{gralla_shape_2020}, in which the accretion disk on the equatorial plane is assumed to be stationary, axisymmetric, and geometrically thin, and each particle in the disk moves along timelike geodesics. similar models were first used for Schwarzschild black holes with stationary emitters \cite{luminet_image_1979}.
In this part, we introduce the thin accretion disk model applicable to Kerr-de Sitter spacetimes.

In this model, the accretion disk is divided into two parts: the part with stable circular orbits and the part with infalling orbits.
For the former,  \(r_{\text{ISCO}}<r<r_{\text{OSCO}}\), 
a particle in the disk has constants of motion \(E_\pm\) and \(L_\pm\) as \autoref{EandLofE} and \autoref{EandLofL}. Using \autoref{eq.geodesic}, 
we obtain the particle's four-velocity as
\begin{equation}
    \bm{u}=u^t\left(\partial _t+\frac{ \pm \sqrt{\frac{1}{r^3}-y} }{\left(1 \pm a \sqrt{\frac{1}{r^3}-y}\right) }\partial _{\phi }\right) \,, \qquad
    u^t=\frac{\Xi \left(1 \pm a\sqrt{\frac{1}{r^3}-y}\right)}{\sqrt{1-\frac{3}{r} \pm 2 a \sqrt{\frac{1}{r^3}-y}-a^2 y}} \,,
    \label{eq:circularU}
\end{equation}
where $\pm$ correspond to the co-rotating orbits and the counter-rotating orbits, respectively. For the latter,  \(r<r_{\text{ISCO}}\), 
we assume that the particle flows along the equatorial geodesic towards the event horizon with the same conserved quantities as that of ISCO:
\begin{equation}
\begin{aligned}
        &u_{\mu }=\left(-E_{\text{ISCO}},u_r,0,L_{\text{ISCO}}\right)     \,.
\end{aligned}
\end{equation}
Together with the condition \(g^{\mu \nu }u_{\mu }u_{\nu }=-1\), we have
\begin{equation}
    u_r=- \sqrt{-\frac{E_{\text{ISCO}}^2g^{tt}-2E_{\text{ISCO}} L_{\text{ISCO}} g^{t\phi}+L_{\text{ISCO}}^2g^{\phi\phi}+1}{g^{rr}}} \,.
\end{equation}

Note that the accretion disk is not specified to be prograde or retrograde here, although the black holes with prograde accretion disk are the most frequently discussed in studies of black hole image. As shown in \autoref{2.2}, the spacetime allows the existence of stable counter-rotating circular orbits,
hence it cannot be ruled out that some astrophysical activity in the cosmos would create black holes with a stable retrograde accretion disk.
In \autoref{5}, we will simulate the images of black holes with prograde and retrograde accretion disks, respectively.

\section{Lensed images with thin disk model}
\label{3}
In this section, we present a comprehensive analytical approach for generating images of Kerr-de Sitter black holes. Firstly, the observer's screen is defined.
Subsequently, a connection is established between each point on the observer's screen and the null geodesic parameterized by two conserved quantities \((\lambda, \eta)\).
Finally, an equation is given to calculate the intensity of each point on the screen.
\subsection[short]{The observer's screen}
\label{3.1}

Firstly, we define the specific spacetime region within which our analysis is conducted. In the case of reasonable parameter values \((\Lambda, a, M)\), the condition \(\Delta _r=0\) indicates the existence of four coordinate singularities known as horizons, namely \(r_{C-}\), \(r_-\), \(r_+\), and \(r_C\) (the cosmological horizon). These singularities act as dividing points, partitioning the spacetime into distinct regions \cite{akcay_kerr-sitter_2011}.
The region of primary interest to us is the interval \([r_+, r_C]\), commonly referred to as the domain of outer communication. In this context, we examine the scenario of an observer free-falling into a black hole while possessing zero angular momentum, also known as Zero Angular Momentum Observer (ZAMO). To describe the observer's local frame, we select an orthonormal tetrad denoted as \(\{\mathbf{e}_{\hat{\mu }}\}\) \cite{Wang:2022kvg}:
\begin{equation}
    \label{tetrad}
    \begin{aligned}
        & \mathbf{e}_{\hat{t}}=\zeta \partial_t+\gamma\partial _{\phi }  \,,\qquad   \mathbf{e}_{\hat{r}}=\frac{1}{\sqrt{g_{r r}}}\partial _r \,,
        \\& \mathbf{e}_{\hat{\theta }}=\frac{1}{\sqrt{g_{\theta  \theta }}}\partial _{\theta }  \,,\qquad  \mathbf{e}_{\hat{\phi }}=\frac{1}{\sqrt{g_{\phi  \phi }}}\partial _{\phi } \,,
        \\& \gamma =-\frac{g_{t \phi } }{g_{\phi  \phi }}\sqrt{\frac{g_{\phi  \phi }}{g_{t \phi }^2-g_{t t} g_{\phi  \phi }}}  \,,\qquad  
        \zeta =\sqrt{\frac{g_{\phi  \phi }}{g_{t \phi }^2-g_{t t} g_{\phi  \phi }}} \,.
    \end{aligned}
\end{equation}

Assuming that any photon reaching the observer's position will be measured by the observer, the general form of the 4-momentum of a photon is given by \autoref{eq.geodesic} with \(\mu=0\). In this context, we use \(k_\mu\) to represent the momentum of the photon. Therefore, the 4-momentum of a photon measured by a ZAMO can be expressed as \(P^{\hat{\mu}}=\eta^{\hat{\mu}\hat{\nu}}\mathrm{e}_{\hat{\nu}}{}^{\xi  }k_{\xi }\), where each component is calculated as follows:
\begin{equation}
    \label{p1234}
    \begin{aligned}
        &P^{\hat{t}}=E(\zeta -\gamma  \lambda )  \,,\qquad   P^{\hat{r}}=E\frac{1}{\sqrt{g_{r r}}}\frac{\pm _r\sqrt{R(r)}}{\Delta_r},
      \\& P^{\hat{\theta }}=E\frac{1}{\sqrt{g_{\theta  \theta }}}\frac{\pm _{\theta }\sqrt{\Theta (\theta )}}{\Delta _{\theta }}  \,,\qquad
      \text{  }P^{\hat{\phi }}=E\frac{\lambda}{\sqrt{g_{\phi\phi }}} \,.
    \end{aligned}
\end{equation}
The relation between the celestial coordinates \(( \vartheta,\varPhi )\) and the local measurement of photons' four-momentum is as follows:
\begin{equation}
    \label{pandab}
    P^{\hat{\phi }}=| P| \sin  \varPhi  \cos  \vartheta \,, P^{\hat{\theta }}=| P| \sin  \vartheta \,, P^{\hat{r}}=| P| \cos  \varPhi  \cos  \vartheta \,, P^{\hat{t}}=|P| \,.
\end{equation}
To display the image on a two-dimensional screen, we project the celestial coordinates onto a plane that faces the black hole \cite{Wang:2022kvg}. The `screen coordinates' \((\alpha ,\beta)\) are related to the celestial coordinates as follows:
\begin{equation}
	\label{abtoxy}
	\alpha=-r_0 \tan  \varPhi \,,\qquad  \beta=r_0 \frac{\tan  \vartheta }{\cos  \varPhi } \,.
\end{equation}
When a photon with four-momentum \(k_\mu\) reaches the observer's position \((t_0,r_0,\theta_0,\phi_0)\), it is displayed on the observer's screen.
By combining the equations \autoref{abtoxy}, \autoref{pandab}, and \autoref{p1234}, we obtain the following relationship:
\begin{equation}
    \label{eqxy}
    \begin{aligned}
            \beta(\lambda ,\eta )&=\left.r\frac{P^{\hat{\theta }}}{P^{\hat{r}}}=\pm _{\theta
            }r\frac{\sqrt{g_{r r}}}{\sqrt{g_{\theta  \theta }}}\frac{\Delta _r\sqrt{\Theta (\theta;\lambda ,\eta  )}}{\Delta _{\theta }\sqrt{R(r;\lambda ,\eta )}}\right| _{\left(r_0,\theta _0\right)} \,,
            \\
            \alpha(\lambda ,\eta )&=\left.-r\frac{P^{\hat{\phi }}}{P^{\hat{r}}}=r\frac{\sqrt{g_{r r}}}{\sqrt{g_{\phi  \phi
            }}}\Delta _r \frac{\lambda }{\sqrt{R(r;\lambda ,\eta )}} \right| _{\left(r_0,\theta _0\right)} \,.
    \end{aligned}
\end{equation}
In fact, \autoref{eqxy} establishes a connection between each point on the observer's screen and a null geodesic parameterized by two conserved quantities \((\lambda,\eta)\). Since spacetime is stationary and axisymmetric, the coordinates \(t_0\) and \(\phi_0\) are unconstrained.
\autoref{eqxy} is a key formula used in constructing the image of black holes. It can be solved inversely for any point \((\alpha ,\beta)\) on the screen to obtain the two conserved quantities \((\lambda,\eta)\) associated with the corresponding null rays. By integrating the geodesic equation, the path of the light ray can be traced backward to infinity or the outer horizon. Fortunately, \autoref{eqxy} is completely invertible, the analytic formulas for \(\lambda(\alpha,\beta)\) and \(\eta(\alpha,\beta)\) are provided in Equations \autoref{eqa:vvv1} and \autoref{eqa:vvv2}, respectively.

\subsection{Observed luminosity}
\label{3.2}
To calculate the observed intensity at each point \((\alpha,\beta)\) on the screen, we need to trace the corresponding light ray backward to infinity or the outer horizon and record each emitting region it passes through. We then add up the emission intensity from all the regions that intersect with the ray. In our case, the only emission region is the accretion disk on the equatorial plane, which is geometrically thin and axisymmetric. Therefore, it is sufficient to record the radius at which the ray intersects the equatorial plane.

After considering the gravitational redshift and neglecting the interaction between the light and the accretion disk, the observed luminosity at a point \((\alpha,\beta)\) on the image plane can be expressed as: \cite{chael_observing_2021}

\begin{equation}
	\label{i=fjg}
	I(\alpha,\beta)=\sum_{m=1}^{m_{\max}} f_m j_{\text{model}}\left(r_m\right)g^3\left(r_m,\alpha,\beta\right) \,,
\end{equation}
where \(r_m\) represents the radius at which the ray intersects the equatorial plane for the \(m^\text{th}\) time, \(m_{\max}\) is the maximum number of equatorial crossings. \(j_{\text{model}}\left(r_m\right)\) denotes the emissivity per unit volume at a specific frequency, and \(g\left(r_m,\alpha,\beta\right)\) represents the redshift factor, which is the observed frequency divided by the emission frequency, at radius \(r_m\). The factor \(g^3\) applies to the intensity of a specific frequency (for instance, 230 GHz image), while \(g^4\) is suitable for integrated intensity \(\int I \dif \nu\) \cite{gralla_black_2019}. \(f_m\) is an optional parameter known as the `fudge factor' that controls the brightness of the higher-order photon ring.

In \cite{chael_observing_2021}, the authors adopt \(f_1=1\) and \(f_{m>1}=2/3\) to best match the time-averaged images for the radiative GRMHD simulation. We follow the same setting for \(f_m\) even though its effect on the overall image is minimal.

In \autoref{2.3}, we divided the accretion disk into two parts with different orbits. Consequently, the redshift factor is expressed as follows:
\begin{equation}
	\label{eq:redshiftfactor}
	g=\frac{\nu _{\text{obs}}}{\nu _{\text{em}}}=\frac{k_{\mu }u_{\text{obs}}^{\mu }}{k_{\mu }u_{\text{em}}^{\mu }}=
	\begin{cases}
		g_\text{circular}(r,\lambda ,\eta ) & r_{\text{ISCO}}<r<r_{\text{OSCO}}  \\
		g_\text{infall}(r,\lambda ,\eta,\nu_r) & r_H<r<r_{\text{ISCO}} \\
	\end{cases} \,,
\end{equation}
and the analytical formulas for \(g_\text{circular}\) and \(g_\text{infall}\) are provided in \autoref{eqa:gin} and \autoref{eqa:gout}.

Based on previous studies investigating black hole accretion disks \cite{abramowicz_foundations_2013,jaroszynski_optics_1997}, the emissivity \(j(r)\) exhibits a significant decrease as the radius \(r\) increases.
In line with the model employed to fit the 1.3 mm (230 GHz) wavelength image of M87* in Chael et al.'s work \cite{chael_observing_2021},
we choose the emissivity as
\begin{equation}
	\text{Log}\left[j_{\text{model}}(r)\right]=-2\text{Log}\left[r/r_+\right]-\frac{1}{2}\text{Log}\left[r/r_+\right]^2 \,,
	\label{eq:emissivity}
\end{equation}
which has also been utilized for simulating images of Kerr-Melvin black holes \cite{hou_image_2022}.

To explicitly simulate the image of Kerr-de Sitter black holes using \autoref{i=fjg}, we still need the specific expressions for \(r_m\) and \(m_{\max }\),
which are uniquely determined by the screen coordinates \((\alpha,\beta)\) through complex analytic ray tracing. In the next section,
we present a complete derivation of \(r_m\) and \(m_{\max }\). By separately addressing the equations for \(r\) and \(\theta\), we obtain \autoref{eq:rm} and \autoref{eq:mmax} as the respective forms of \(r_m\) and \(m_{\max }\).

\section{Analytic ray tracing}
\label{4}
Refs. \cite{gralla_lensing_2020} and \cite{gralla_null_2020}  have provided complete analytic solutions for null geodesics and derived the analytical formulas for \(r_m(\alpha,\beta)\) and \(m_{\max}(\alpha,\beta)\) in Kerr spacetimes. Additionally, in \cite{omwoyo_black_2023}  the authors obtained analytic solutions in terms of elliptic integrals for near-bound null geodesics in Kerr-de Sitter spacetimes. Building upon these previous works, we derive expressions of \(r_m(\alpha,\beta)\) and \(m_{\max}(\alpha,\beta)\) that apply to Kerr-de Sitter spacetimes.

The null geodesic equations for \(r\) and \(\theta\) are derived from \autoref{nullgeos} as follows:
\begin{equation}
    \label{geo-randtheta}
    \frac{\Sigma }{E}p^r=\pm _r\sqrt{R(r)} \,,\qquad \frac{\Sigma }{E}p^{\theta }=\pm _{\theta }\sqrt{\Theta (\theta )}\,,
\end{equation}
where \(R(r)\) and \(\Theta (\theta )\) are given in \autoref{RRandTheta}. 
The common approach to decoupling geodesic equations is to parameterize the geodesics using the `Mino time' \(\tau\)  \cite{mino_perturbative_2003}, defined as:
\begin{equation}
	\frac{d x^{\mu }}{d \tau }=\frac{\Sigma }{E}p^{\mu} \,.
\end{equation}
Using this parameterization, the total Mino time that a photon experiences from the observer's position \((r_0,\theta_0)\) to the emitting source's position \((r_s,\theta_s)\) can be expressed as:
\begin{subequations}
	\begin{align}
		\label{rintegral}
		\int_{r_0}^{r_s}\frac{d r}{ \pm _r\sqrt{R(r)}}&=I_r \,,
		\\ \int_{\theta_0}^{\theta_s} \frac{d \theta }{\pm _{\theta }\sqrt{\Theta (\theta )}}&=G_\theta \,,\label{thetaintegral}
		\\ I_r=G_\theta=&\int d\tau \,,
	\end{align}
\end{subequations}
where the integrals represent integration along the path. The plus or minus signs, \(\pm_r\) and \(\pm_{\theta}\), are chosen depending on the direction of integration.

\subsection[short]{Radial equation}
\label{4.1}
We now turn our attention to the radial equation. Introducing the radial roots \((r_1, r_2, r_3, r_4)\), which satisfy the following relation:
\begin{equation}
    R(r)=\mathcal{E}^2(r-r_1)(r-r_2)(r-r_3)(r-r_4)=0 \,,
\end{equation}
with
\begin{equation}
    \mathcal{E}=\sqrt{\frac{\eta  \Lambda }{3}+\Xi ^2+\frac{1}{3} a^2 \Lambda  \Xi ^2-\frac{2}{3} a \lambda  \Lambda  \Xi ^2+\frac{1}{3} \lambda ^2\Lambda  \Xi ^2} \,.
\end{equation}
These radial roots are determined by the parameters \((\lambda, \eta)\) of the null geodesics, and their analytical solutions can be obtained by Ferrari's method \cite{omwoyo_black_2023}.
We are particularly interested in the geodesics near the black hole, which can be classified into two groups based on their associated radial roots: 

1. Geodesics with four real roots \((r_1 < r_2 < r_3 < r_4)\), where \(r_0 > r_4 > r_+\). These geodesics have one turning point and extend to infinity. 

2. Geodesics that never encounter a turning point and eventually cross the event horizon. 

Therefore, the total Mino time \(I_r^{\text{total}}\) required for light rays from the observer to reach either infinity or the horizon can be expressed as:
\begin{equation}
    I_r^{\text{total}}=
    \begin{cases}
    -\int _{r_0}^{r_4}\frac{d r}{\sqrt{R(r)}}+\int _{r_4}^{\infty }\frac{d r}{\sqrt{R(r)}} & r_0>r_4>r_+ \,,
    \\-\int _{r_0}^{r_+}\frac{d r}{\sqrt{R(r)}}  &  r_4\in  \mathbb{C} \,\text{or}\, r_4<r_+  \,.
    \end{cases}
\end{equation}
To obtain the expression for \(r(\tau)\), the radial equation need to be solved for different cases.
\footnote{For detailed calculations, please refer to  \cite{gralla_null_2020} or \cite{omwoyo_black_2023}.}

For the cases where \(r_4\in\mathbb{R}\), solving \autoref{rintegral} in reverse provides the solution:
\begin{equation}
    \label{eq:rIr1}
    r(I_r )=\frac{r_3r_{14}\text{sn}^2\left(\mathit{g}_E I_r +\nu _rF\left(\phi _{E,0}|k_E\right)|k_E\right)+r_4r_{31}}
    {r_{14}\text{sn}^2\left(\mathit{g}_E I_r+\nu _rF\left(\phi _{E,0}|k_E\right)|k_E\right)+r_{31}}   \qquad  (r_4\in\mathbb{R})\,,
\end{equation}
where \(\text{sn}(\phi|k)\) is the Jacobi elliptic sine function \cite{abramowitz1948handbook}, \(F(\phi|k)\) is the elliptic integral of the first kind, and
\begin{equation}
	\begin{aligned}
		& r_{ij}=r_i-r_j \,,\qquad k_E=\frac{r_{41}r_{32}}{r_{42}r_{31}} \,,
		\\&  \mathit{g}_E=\frac{\mathcal{E}\sqrt{r_{42}r_{31}}}{2}  \,,\qquad   \phi _{E,i}=\arcsin \left(\sqrt{\frac{r_{31}r_{i4}}{r_{41}r_{i 3}}}\right) \,.
	\end{aligned} 
\end{equation}

For the cases where \(r_4=\bar{r}_3\in\mathbb{C}\), the inverse solution of \autoref{rintegral} yields
\begin{align}
	\label{eq:rIr2}
	r(I_r)=\frac{\left(r_2B+A r_1\right)\text{cn}\left(\mathcal{E}\sqrt{A B} I_r +\nu _rF\left(\phi _{P,0}|k_P\right)|k_P\right)+\left(r_2B-A r_1\right)}
	{(B+A)\text{cn}\left(\mathcal{E}\sqrt{AB} I_r +\nu _rF\left(\phi _{P,0}|k_P\right)|k_P\right)+(B-A)}   \qquad (r_4=\bar{r}_3\in\mathbb{C})  \,,
\end{align}
where \(\text{cn}(\phi|k)\) is the Jacobi elliptic cosine function \cite{abramowitz1948handbook}, and
\begin{equation}
	\begin{aligned}
        & A=\sqrt{r_{32}r_{42}} \,,\qquad B=\sqrt{r_{31}r_{41}}  \,,\\
		& k_P=\frac{(B+A)^2-\left(r_2-r_1\right){}^2}{4A B} \,,\qquad \phi _{P,i}=\arccos \left(\frac{(A-B)r_i+r_2B-r_1A}{(A+B)r_i-r_2B-r_1A}\right) \,.
	\end{aligned}
\end{equation}
The independent variable \(I_r\) in  \autoref{eq:rIr1} or \autoref{eq:rIr2} is limited to the range \(0<I_r<I_r^\text{total}\), where
\begin{align}
    \label{eq:Itotal}
    I_r^\text{total}=
    \begin{cases}
        -\mathit{g}_E\left[F\left(\left.\phi _{E,4}\right|k_E\right)-F\left(\left.\phi _{E,0}\right|k_E\right)\right]+\mathit{g}_E\left[F\left(\left.\phi
        _{E,\infty }\right|k_E\right)-F\left(\left.\phi _{E,4}\right|k_E\right)\right]  &  r_4>r_+ \in\mathbb{R} \,,
        \\-\mathit{g}_E\left[F\left(\left.\phi _{E,+}\right|k_E\right)-F\left(\left.\phi _{E,0}\right|k_E\right)\right]  &  r_4<r_+ \in\mathbb{R} \,,
        \\-\frac{1}{\mathcal{E}\sqrt{A B}}\left[F\left(\phi _{P,+}|k_P\right)-F\left(\phi_{P,0}|k_P\right)\right]  &  r_4=\bar{r}_3 \in\mathbb{C} \,.
    \end{cases}
\end{align}

\autoref{eq:rIr1} and \autoref{eq:rIr2} together constitute the solution of the radial equation in  the domain of outer communication,
which will be used to calculate the radius \(r_m\) of the photon when it crosses the equatorial plane for the \(m^\text{th}\) time.

\subsection[short]{Angular integrals}
\label{4.2}
In this part, we obtain the Mino time experienced by photons when they cross the equatorial plane by addressing the angular integral.

Letting \(u=\cos^2\theta\), the angular potential \(\Theta(\theta)\) can be written as
\begin{align}
    \Theta (\theta )&=\frac{1}{1-u}\mho \left(u-u_+\right)\left(u-u_-\right) \,,
    \\\mho &=-\frac{1}{3} a^2 \eta  \Lambda -a^2 \Xi ^2-\frac{1}{3} a^4 \Lambda  \Xi ^2+\frac{2}{3} a^3 \lambda \Lambda \Xi ^2-\frac{1}{3} a^2 \lambda^2 \Lambda \Xi ^2 \,,
\end{align}
with
\begin{equation}
    \begin{aligned}
        u_{\pm }&=\Delta \pm \sqrt{\frac{3 \eta }{a^2 \left(\Lambda  \left(\Xi ^2 (a-\lambda )^2+\eta \right)+3 \Xi ^2\right)}+\Delta ^2} \,,
        \\\Delta &=\frac{\eta  \left(a^2 \Lambda -3\right)+\Xi ^2 (a-\lambda ) \left(a^3 \Lambda -a^2 \lambda  \Lambda +3 a+3 \lambda \right)}{2 a^2 \left(\Lambda  \left(\Xi ^2 (a-\lambda )^2+\eta \right)+3 \Xi ^2\right)} \,.
    \end{aligned}
\end{equation}
As demonstrated in \cite{omwoyo_black_2023}, a photon following ordinary geodesics (\(\eta>0\)) undergoes libration between \(\theta_+=\arccos(\sqrt{u_+})\) and \(\theta_-=\arccos(-\sqrt{u_+})\), crossing the equatorial plane each time. Consequently, the path integral in  \autoref{thetaintegral} can be expanded and expressed as:
\begin{equation}
    \begin{aligned}
        G_{\theta }\left(\theta _{s}\right)=2m\left|\int _{\pi /2}^{\theta _{\pm }}\right|\frac{d \theta }{\sqrt{\Theta (\theta )}}+
        \eta _0\left|\int_{\pi /2}^{\theta _0}\right|\frac{d \theta }{\sqrt{\Theta (\theta )}}-
        \eta _{s }\left|\int _{\pi /2}^{\theta _{s }}\right|\frac{d\theta }{\sqrt{\Theta (\theta )}} \,,
    \end{aligned}
\end{equation}
where \(m\) represents the number of times the ray reaches the equatorial plane, and
\begin{equation}
    \left.\eta _{i }=\text{sign}(p_\theta\cos\theta)\right|_{\theta=\theta_i} \,.
\end{equation}
After reducing the integral to the form of elliptic integrals, we can obtain the Mino time it takes for the ray to reach the equatorial plane for the \(m^\text{th}\) time as follows:
\begin{align}
    \label{eq:Gm}
    G_m=\frac{2\left(m-H(\eta _{0})\right)}{\sqrt{-u_-\mho}}K\left(\frac{u_+}{u_-}\right)+
    \frac{\eta_{0 }}{\sqrt{-u_-\mho}}F\left(x_0\left|\frac{u_+}{u_-}\right.\right) \,,
\end{align}
where \(H\) denotes the Heaviside function \cite{abramowitz1948handbook}, and
\begin{align}
    x_0=\arcsin \sqrt{\frac{\cos ^2\theta _{0}}{u_+}} \,.
\end{align}
By utilizing \autoref{eq:rIr1} or \autoref{eq:rIr2}, we can determine the radius \(r_m(\alpha,\beta)\) at which the ray crosses the equatorial plane for the \(m^\text{th}\) time as follows:
\begin{align}
    r_m(\alpha ,\beta )=r\left(G_m\right) \,,
    \label{eq:rm}
\end{align}
while \(0<G_m<I_r^\text{total}\), and the maximal value of \(m\) is
\begin{align}
    \label{eq:mmax}
    m_{\max }=
    \left[\frac{\sqrt{-u_-\mho}I_r^{\text{total}}-\eta _{\theta }F\left(x_0\left|\frac{u_+}{u_-}\right.\right)}{2K\left(\frac{u_+}{u_-}\right)}-H(\eta _{0})\right] \,,
\end{align}
where \(I_r^\text{total}\) is given in \autoref{eq:Itotal}, and the bracket indicates the rounding down operation.

\begin{figure}[ht]
    \centering
    \includegraphics[width=\textwidth]{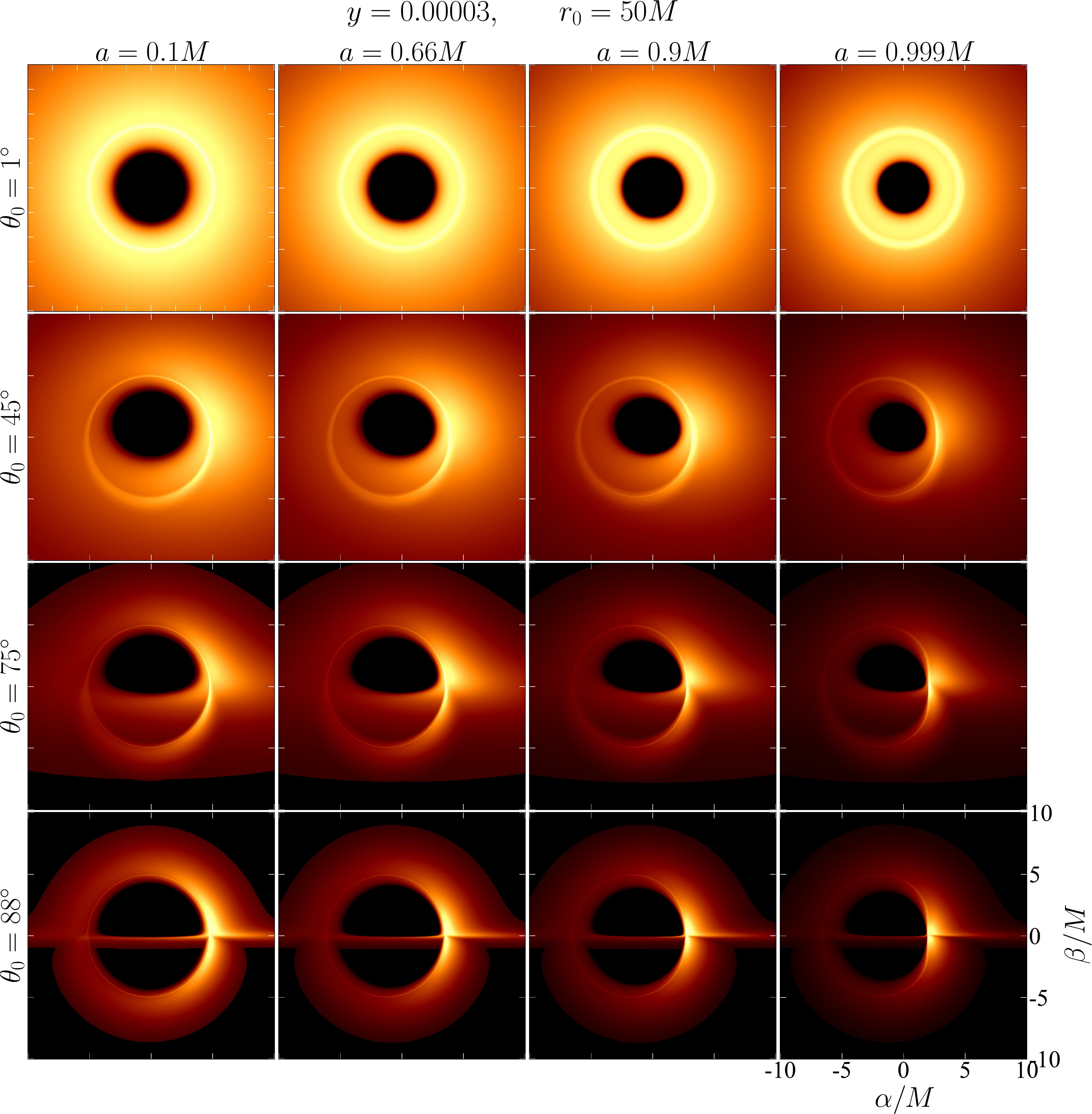}

    \caption{The images depict Kerr-de Sitter black holes illuminated by a prograde thin equatorial accretion disk. In these simulations, the dimensionless cosmological constant is fixed at \(y=0.00003\), and the radial observation distance \(r_0\) is consistently set at \(50M\) from the black hole. The columns, from left to right, represent models with spins \(a=(0.1, 0.66, 0.9, 0.999)\), while the rows, from top to bottom, display observations at inclinations \(\theta_0=(1^\circ, 45^\circ, 75^\circ, 88^\circ)\). For each graph, the intensity data is independently normalized and then mapped to colors using the colormap 'afmhot`, employing a power function with an index of \(\gamma=1/4\).}
    \label{fig:grid-prog}
\end{figure}

\begin{figure}[ht]
    \centering
    \includegraphics[width=\textwidth]{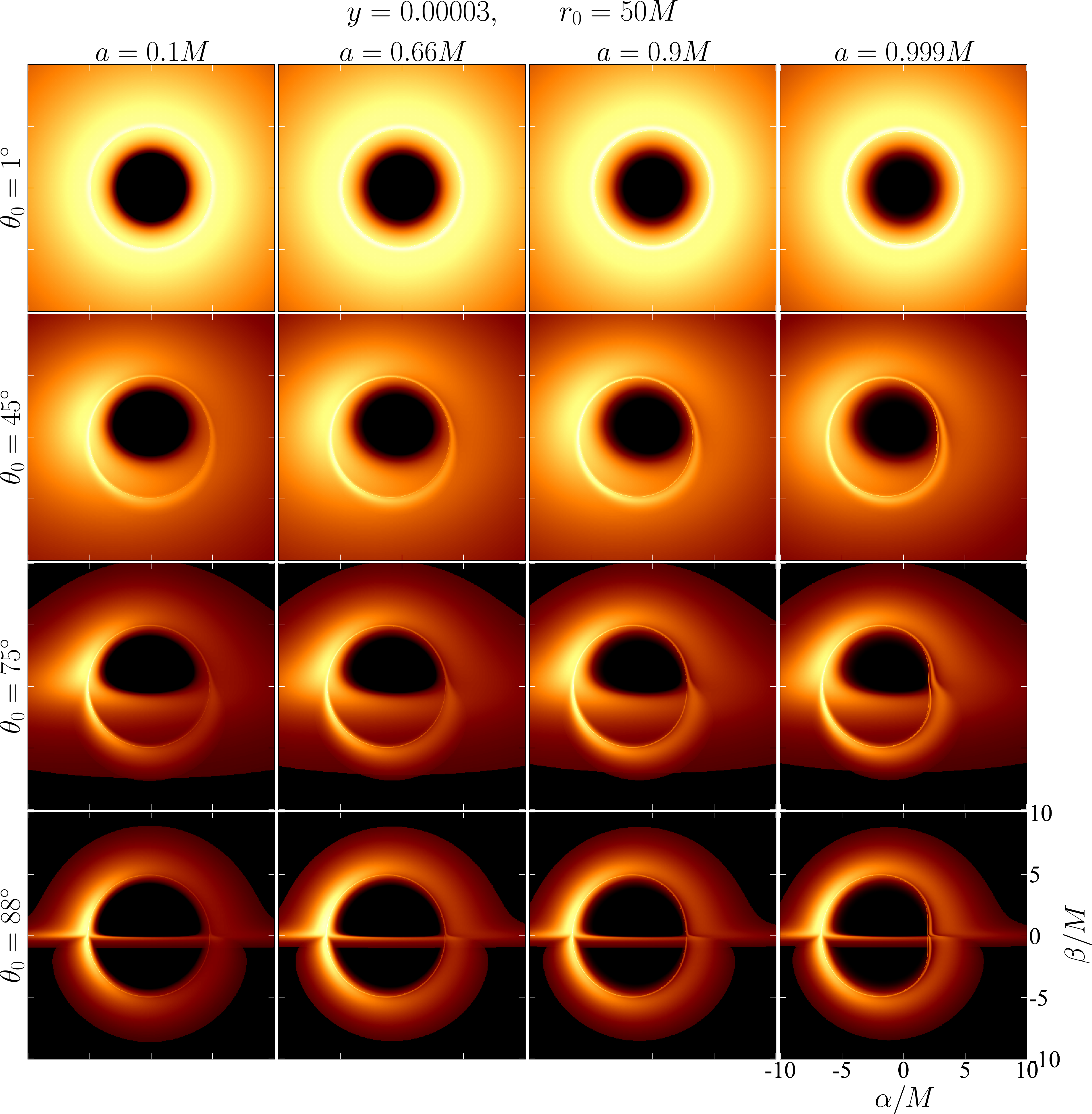}

    \caption{The images portray Kerr-de Sitter black holes illuminated by a retrograde thin equatorial accretion disk. In these simulations, the dimensionless cosmological constant is fixed at \(y=0.00003\), and the radial observation distance \(r_0\) remains consistently \(50M\) away from the black hole. The columns, from left to right, represent models with spins \(a=(0.1, 0.66, 0.9, 0.999)M\), while the rows, from top to bottom, display observations at inclinations \(\theta_0=(1^\circ, 45^\circ, 75^\circ, 88^\circ)\). For each graph, the intensity data is independently normalized and mapped to colors using the color map 'afmhot`, employing a power function with an index of \(\gamma=1/4\).}
    \label{fig:grid-retrog}
\end{figure}

\section{Simulations}
\label{5}
The mathematical preparations for simulating images of Kerr-de Sitter black holes have now been completed. In this section, we explore the potential effects of the cosmological constant on black hole observations through various simulations. These simulations encompass the overall visual appearance of the black hole images, the characteristics of the observed curves, the relative size of the images, and the observed intensity. By conducting these simulations, we aim to gain insights into the impact of the cosmological constant on the observations of black holes.

\subsection[short]{Images with prograde and retrograde disk}
\label{5.1}
We use \autoref{i=fjg} to generate simulated images of Kerr-de Sitter black holes. Recall that our model is a generalization of the M87* analytic model developed by \cite{chael_observing_2021}. While \cite{chael_observing_2021} considered a static observer at infinity and utilized broken power-law fitting functions to represent the 4-velocity of accreting gas, we consider a ZAMO and describe the 4-velocity of the accreting gas using precise time-like orbits in Kerr-de Sitter spacetimes. Additionally, our choices of the emissivity in  \autoref{eq:emissivity} and the `fudge' factor in  \autoref{i=fjg} align with those of \cite{chael_observing_2021}.

In \autoref{fig:grid-prog} and \autoref{fig:grid-retrog}, we present simulations of the observations of Kerr-de Sitter black holes with prograde \autoref{fig:grid-prog} and retrograde \autoref{fig:grid-retrog} accretion disks, respectively. We set the dimensionless cosmological constant to \(y=0.00003\) for both cases. For different values of the black hole spin \(a=(0.1, 0.66, 0.9, 0.999)M\), we examine the observation inclination angles \(\theta_0=(1^\circ, 45^\circ, 75^\circ, 88^\circ)\), respectively. The observation distance remains fixed at \(r_0=50M\), and the observer's screen range is restricted to \(-10<\alpha/M<10\) and \(-10<\beta/M<10\).

As demonstrated in our simulated images, the presence of the OSCO in Kerr-de Sitter spacetimes leads to the clear visibility of the outer boundary of the accretion disk when the observer inclination is close to 90 degrees. Each image exhibits a distinct bright ring, known as the `photon ring', which closely aligns with the critical curve. The dark regions at the center correspond to the event horizon of the black hole. Overall, the presence of a cosmological constant has minimal impact on the visual appearance of the black hole.

\subsection[short]{The characteristic curves}
\label{5.2}
To analyze the geometric details of black hole images, a straightforward approach is to plot the various geometric boundaries of the model directly on the observer's screen. These boundaries represent the apparent positions of the geometric features. A light source can have an infinite number of apparent positions because the emitted ray can travel around the black hole any number of times before reaching the observer. Each apparent position is labeled with a number \(m=1, 2, 3, \ldots\), where even \(m\) corresponds to the back of the equatorial disk and odd \(m\) corresponds to the front of the disk.

For the entire equatorial emission surface, its image on the observer's screen appears as infinitely many rings, known as photon rings \cite{johnson_universal_2020,gralla_lensing_2020}. Each photon ring represents a miniature image of the entire equatorial plane. Consequently, the \(m^\text{th}\) order image of a curve on the equatorial plane is confined to the \(m^\text{th}\) order photon ring. As \(m\) approaches infinity, the width of the photon rings becomes negligible, allowing us to describe the photon ring as a closed curve \(\mathcal{C}\), commonly known as the photon's `critical curve'.

The photon's critical curve \(\mathcal{C}\) has been extensively studied in previous works \cite{grenzebach_photon_2014,li_shadow_2020,omwoyo_remarks_2022}. In terms of screen coordinates, using \autoref{eqxy}, the curve \(\mathcal{C}\) can be expressed as:
\begin{equation}
    \vec{\mathcal{C}} (R)=(\alpha(\lambda (R),\eta (R)),\pm \beta(\lambda (R),\eta (R))) \qquad R\in[r^\gamma_{+},r^\gamma_-] \,,
\end{equation}
where \(\lambda (R)\), \(\eta (R)\) and \(r^\gamma_{\pm}\) are derived in \autoref{appendix:critical-curve}.
In addition to the critical curve, we are also interested in other characteristic curves defined by a constant Boyer-Lindquist radius \(r=\text{const.}\), particularly the equatorial horizon and the ISCO. These curves, in screen coordinates, can be represented by the equation
\begin{align}
	r_m(\alpha,\beta) = \text{const.} \,,
\end{align}
where \(r_m(\alpha,\beta)\) is defined in \autoref{eq:rm}.

\begin{figure}
    \centering
    \includegraphics[width=0.32\textwidth]{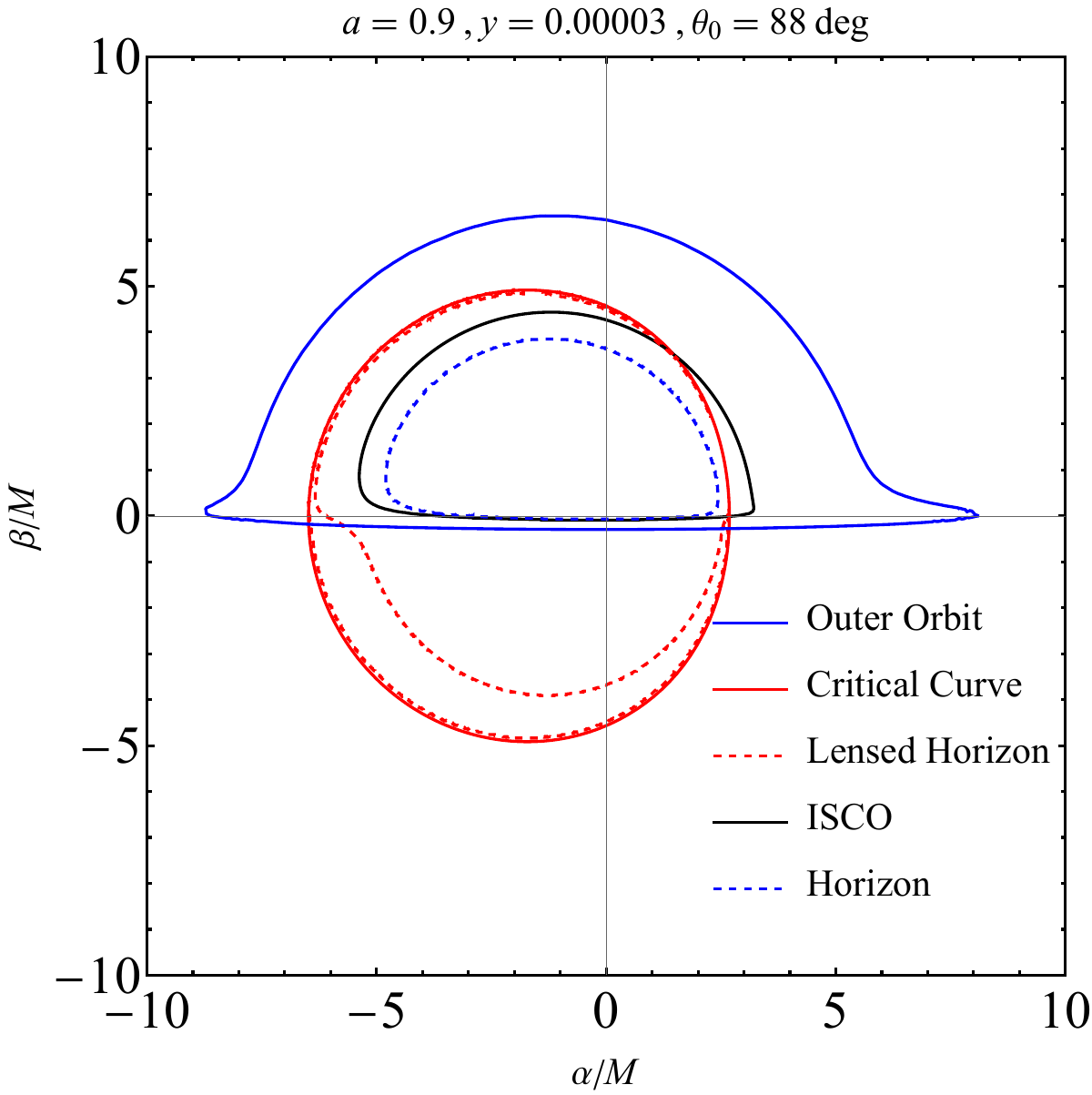}
    \includegraphics[width=0.32\textwidth]{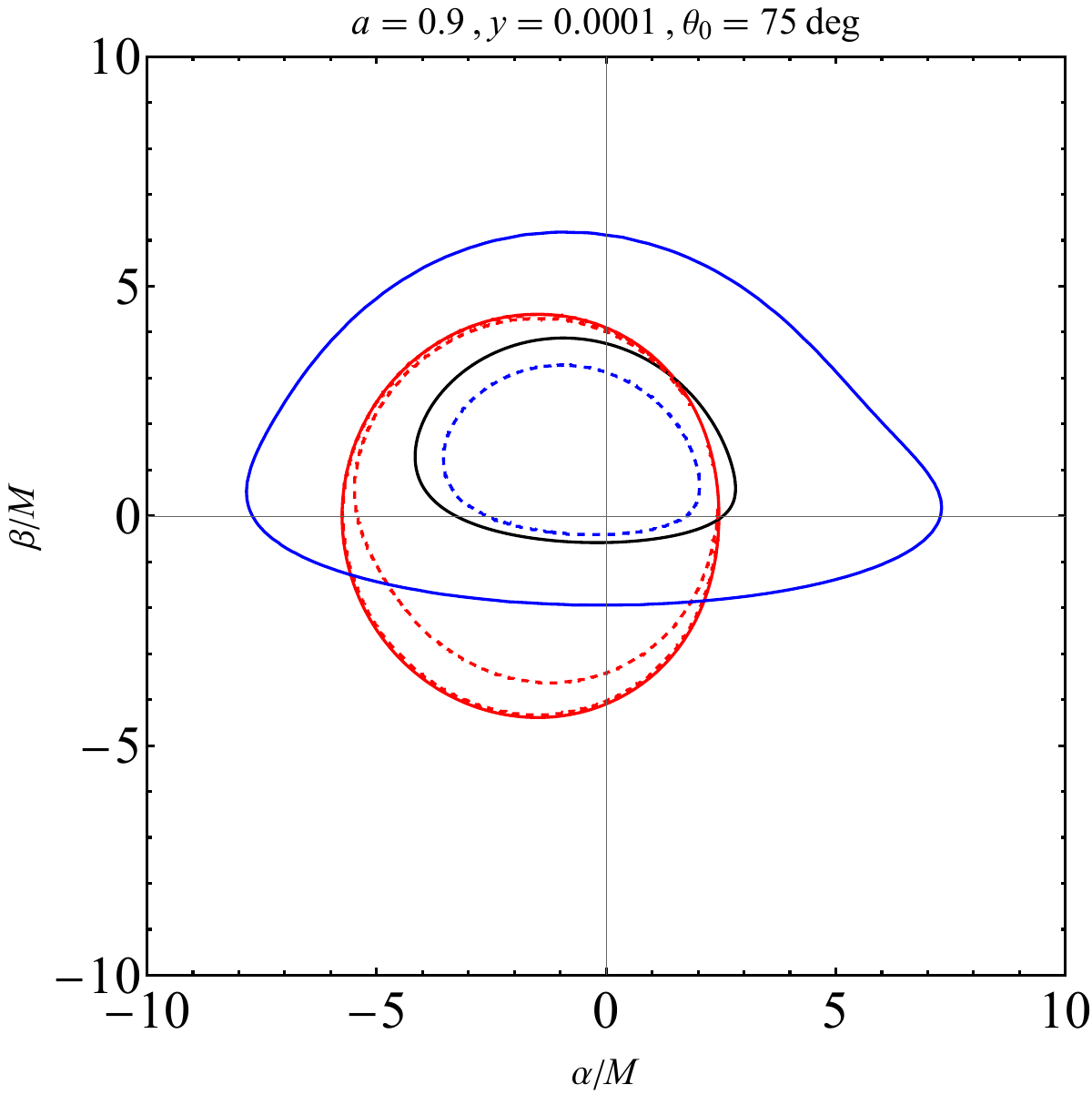}
    \includegraphics[width=0.32\textwidth]{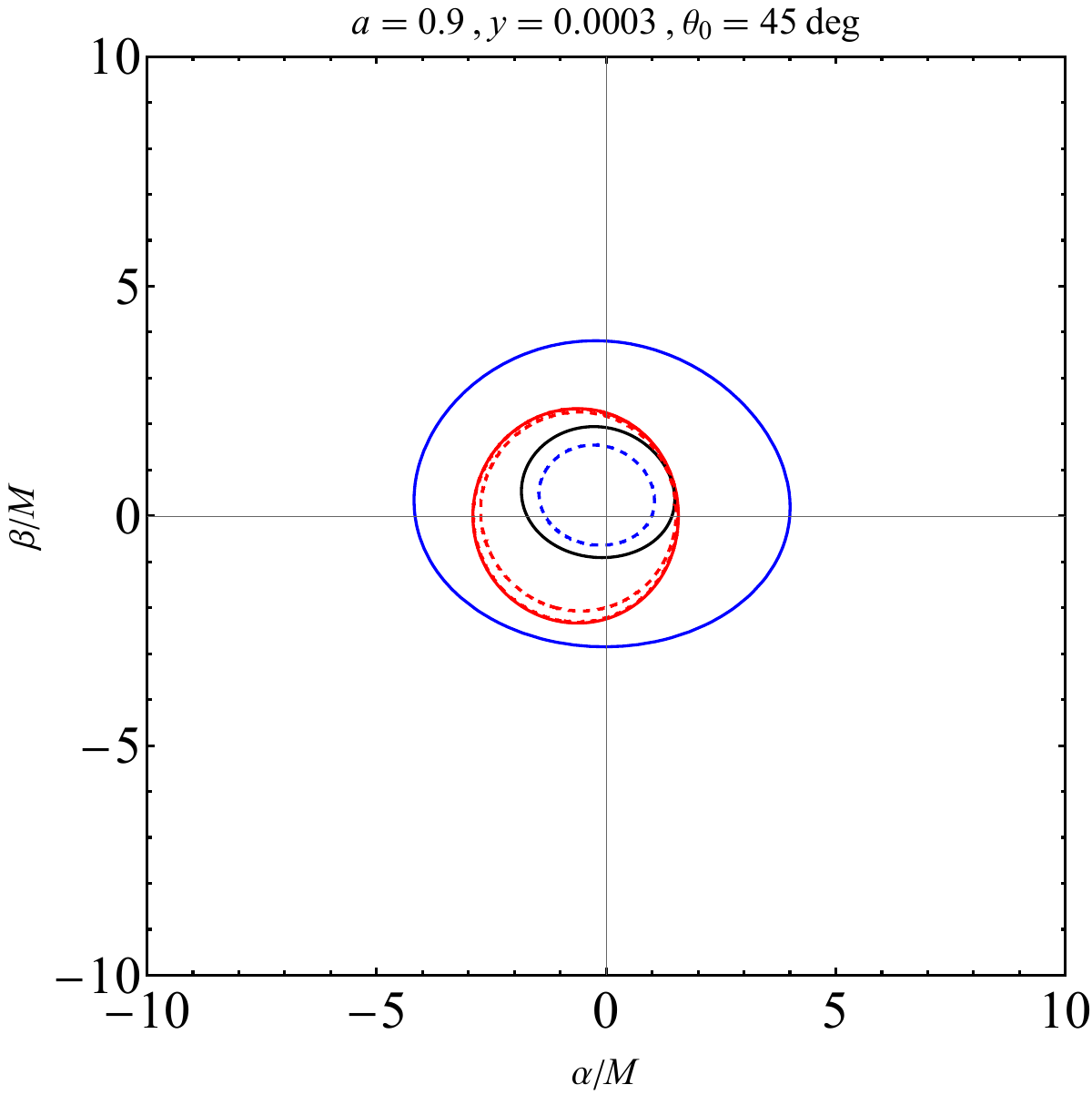}
    \includegraphics[width=0.32\textwidth]{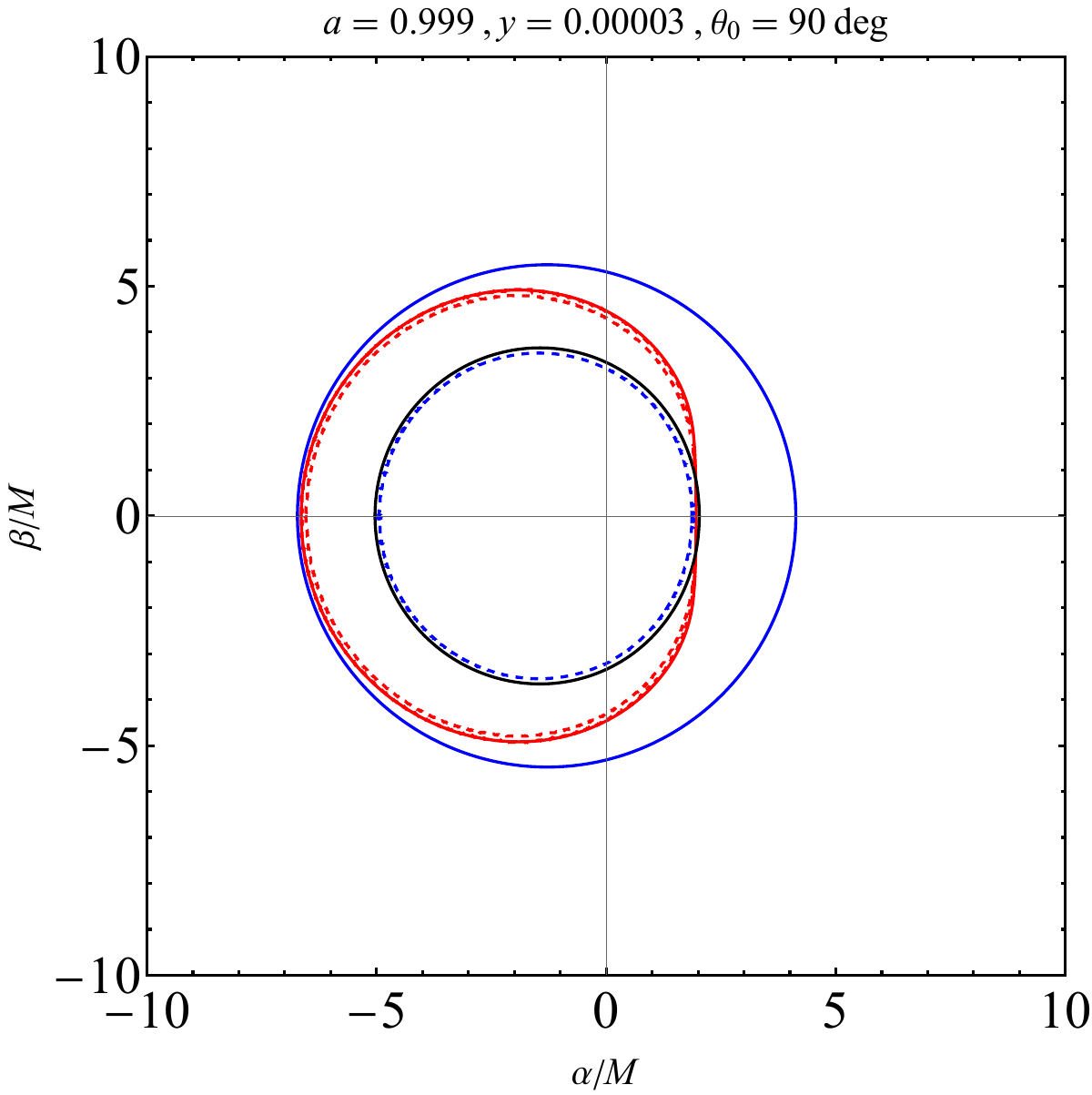}
    \includegraphics[width=0.32\textwidth]{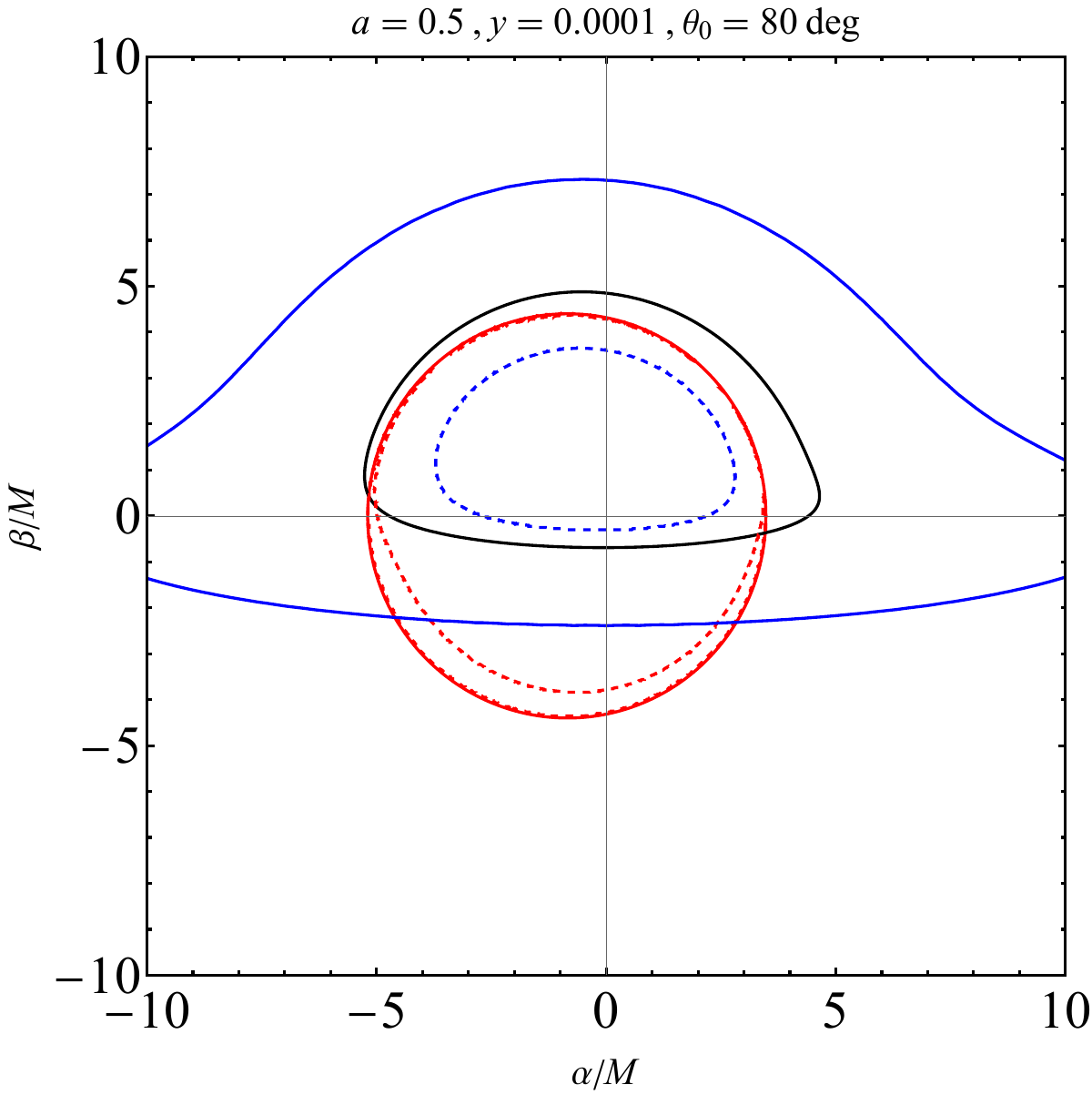}

    \caption{The characteristic curves depicted in the figure depend on the black hole spin, cosmological constant, and observer inclination. The red solid curves represent the critical curve of photons. The blue solid lines correspond to the direct image (\(m=1\)) of the equatorial circular orbit with a constant Boyer-Lindquist radius \(r=2(r_H+r_{\text{ISCO}})\). The black solid lines represent the direct image (\(m=1\)) of the ISCO. The red dashed lines represent the lensed image (\(m=2\)) of the equatorial horizon. The blue dashed lines depict the direct image (\(m=1\)) of the equatorial horizon. The specific black hole parameters and observer inclination are indicated at the top of each panel.}
    \label{fig:curves}
\end{figure}

In \autoref{fig:curves}, we present several characteristic curves of different black hole models. These include the direct image (\(m=1\)) of the equatorial horizon, the ISCO, and the circular orbit at \(r=2(r_H+r_{\text{ISCO}})\). We also show the lensed image (\(m=2\)) of the equatorial horizon and the critical curve.

\subsection{The relative size}
\label{5.3}

\begin{figure}[htbp]
    \centering
    \includegraphics[width=\textwidth]{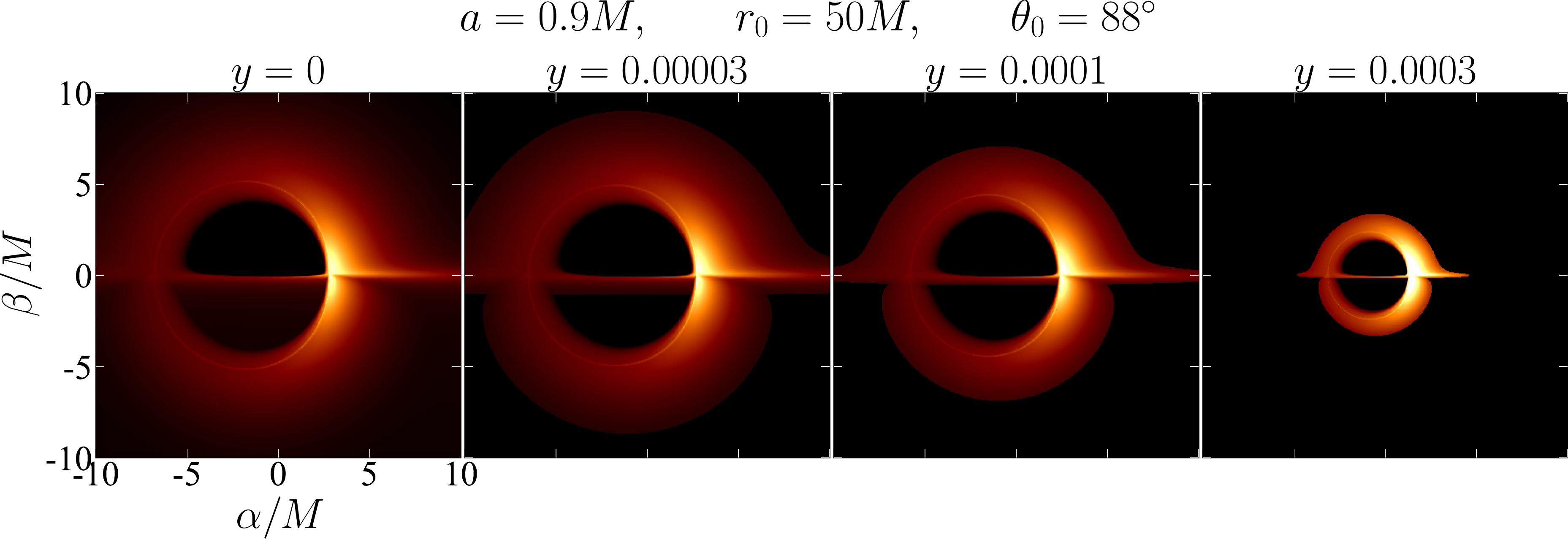}

    \caption{Images of Kerr-de Sitter black holes with different cosmological constants. From left to right, the dimensionless cosmological constants are set as
    \(y=(0,0.00003,0.0001,0.0003)\). Each black hole in the figure has the same spin \(a=0.9\) and the observer's positions are fixed at \((r_0=50M,\theta_0=88^\circ)\).}
    \label{fig:bigtosmall}
\end{figure}

Previous studies on the black hole shadow have revealed that the cosmological constant leads to a reduction in the size of the shadow \cite{grenzebach_photon_2014,li_shadow_2020}. As depicted in  \autoref{fig:curves}, the other characteristic curves of the black hole also diminish in size in proportion to the critical curve. Consequently, it is evident that the overall black hole image will shrink accordingly.
To illustrate this phenomenon, we showcase \autoref{fig:bigtosmall}, featuring simulated images of black holes at varying cosmological constants.

While previous discussions concentrated on a fixed observation distance \(r_0\), the presence of a cosmological constant introduces a notable influence on the relative size of the black hole shadow and image, which is further contingent on the observation distance. To scrutinize the correlation between the relative size of the black hole, the cosmological constant, and the observation distance, we introduce a parameter \(\rho_w\) as a representation of the black hole's relative size.



\begin{figure}[htbp]
    \centering
    \includegraphics[width=0.495\textwidth]{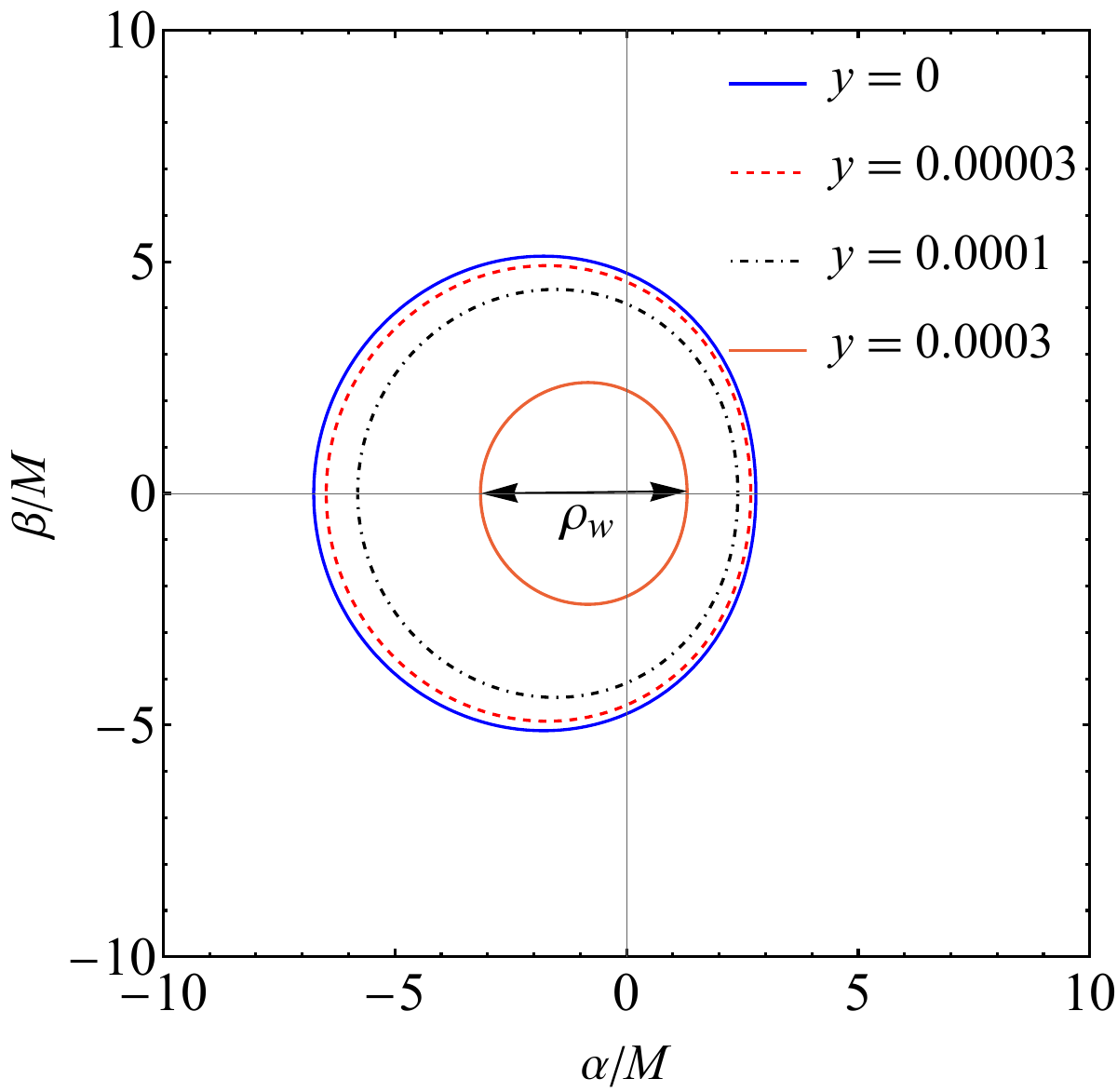}
    \caption{The critical curves of the four black holes in \autoref{fig:bigtosmall}. 
    The curves from outside to inside possess dimensionless cosmological constants \(y=0\) (blue solid), 
    \(y=0.00003\) (red dasded), \(y=0.0001\) (blue dashed), and \(y=0.0003\) (red solid).}
    \label{fig:four-shadow}
\end{figure}

For simplicity, we will adopt a fixed observation inclination (\(\theta_0=90^\circ\)) and black hole spin (\(a=0.9M\)), exclusively directing our attention to the discussion of the black hole's relative size. In the context of this analysis, the parameter \(\rho_w\) is defined as the horizontal radius of the black hole shadow, as illustrated in \autoref{fig:four-shadow}. Consequently, \(\rho_w\) is concisely expressed as:
\begin{equation}
    \label{eq:rho}
    \rho_w=\alpha(\lambda (r^\gamma_+),\eta (r^\gamma_+))-\alpha(\lambda (r^\gamma_-),\eta (r^\gamma_-)) \,,
\end{equation}
where \(\alpha(\lambda ,\eta)\) is given in \autoref{eqxy}, \(\lambda (R)\), \(\eta (R)\) and \(r^\gamma_\pm\) are given in \autoref{appendix:critical-curve}. 

In \autoref{fig:shadowsize}, the demonstration reveals that the relative size \(\rho_w\) of the black hole diminishes with increasing observation distance \(r_0\) or cosmological constant. Noteworthy is the consistent reduction of the relative size to zero as the limit \(r_0 \to r_C\) is approached, where \(r_C\) signifies the cosmological horizon introduced at the outset of \autoref{3.1}. This observation prompts an investigation into the correlation between the relative size and the relative observation position \(r_0/r_C\).

\begin{figure}[htbp]
    \centering
    \includegraphics[width=0.495\textwidth]{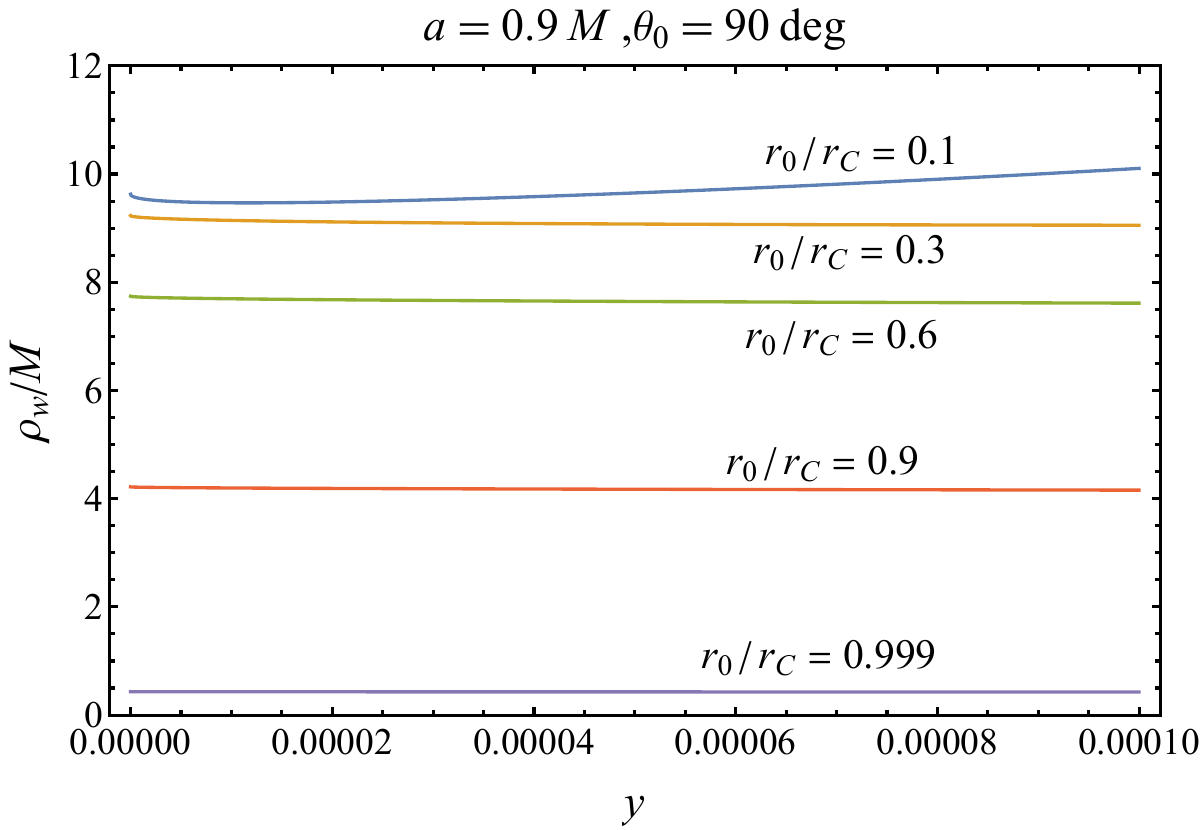}
    \includegraphics[width=0.495\textwidth]{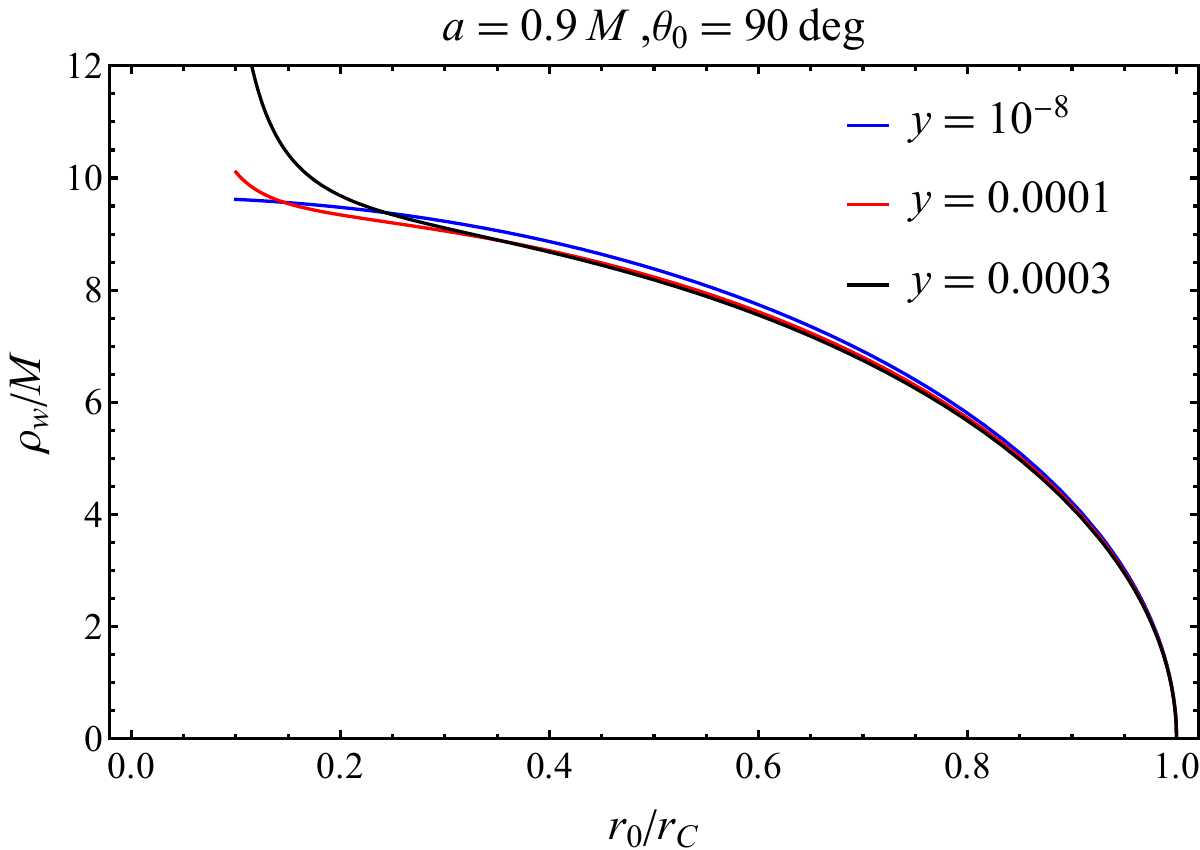}
    \caption{The relative size of the shadow varies with respect to \(y\) (the dimensionless cosmological constant) or 
    \(r/r_C\) (the relative observation distance)}
    \label{fig:newsize}
\end{figure}

As depicted in \autoref{fig:newsize}, the relative size of the black hole remains largely constant when the relative observation position is fixed, showing minimal influence from the cosmological constant. Nevertheless, with a non-vanishing cosmological constant, the relative size of the black hole consistently decreases as the observer moves farther away from it.

\begin{figure}[htbp]
    \centering
    \includegraphics[width=0.495\textwidth]{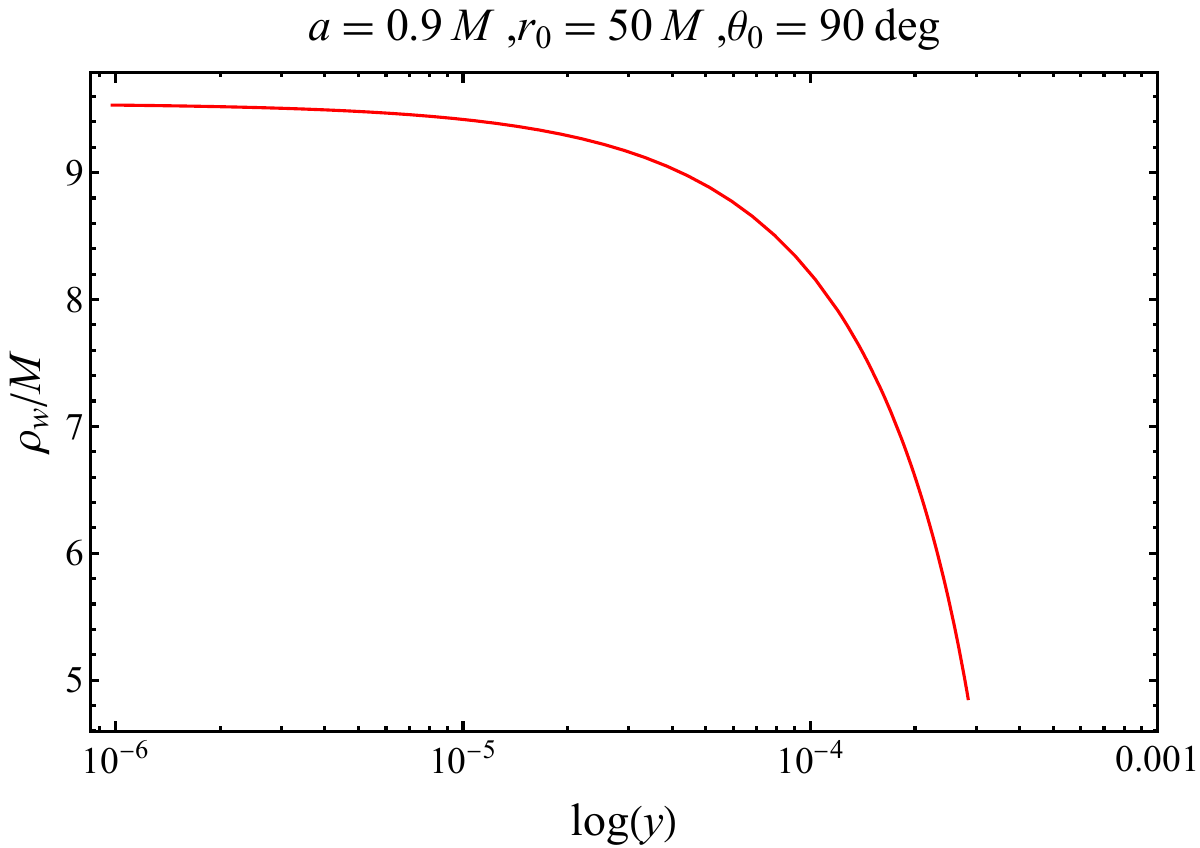}
    \includegraphics[width=0.495\textwidth]{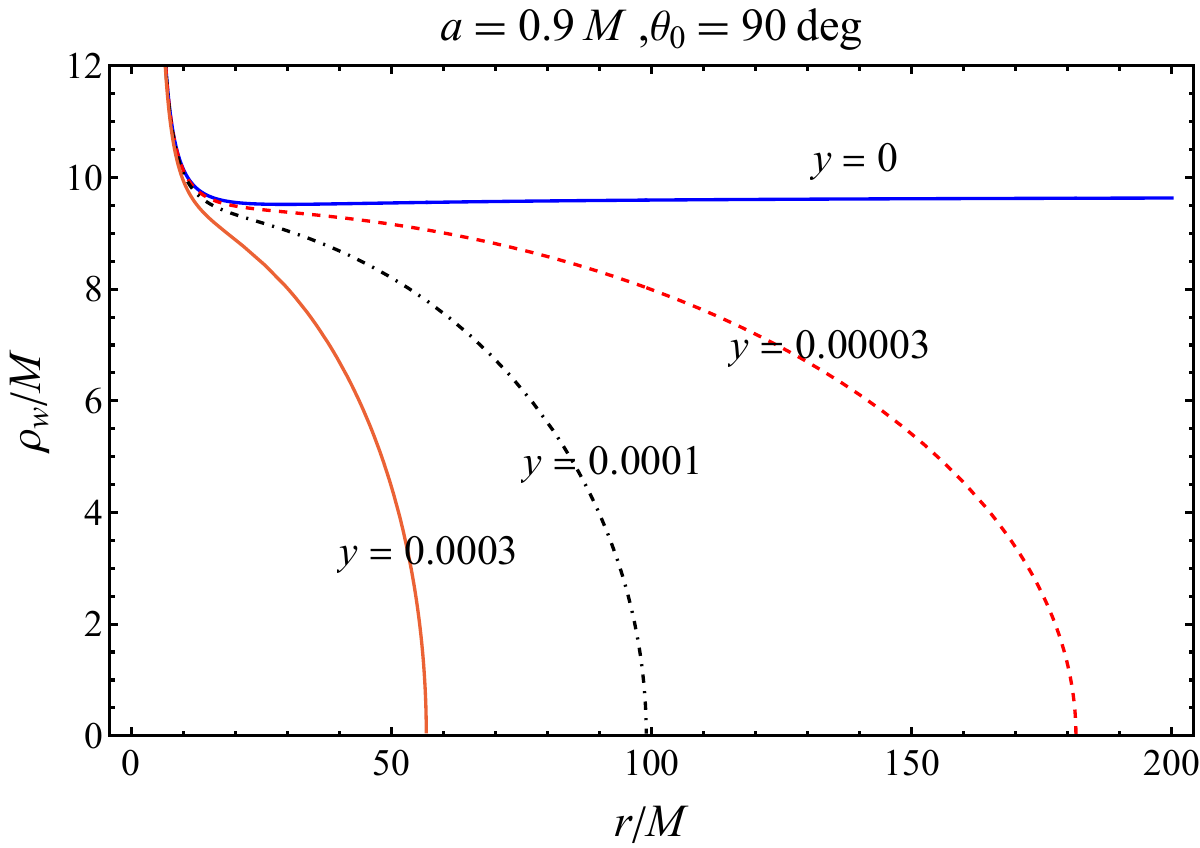}
    \caption{The relative size of the shadow varies with respect to \(y\) (the dimensionless cosmological constant) or 
    \(r\) (the observation distance).}
    \label{fig:shadowsize}
\end{figure}

\subsection{Observed intensity}
\label{5.4}
\begin{figure}[htbp]
    \centering
    \includegraphics[width=0.495\textwidth]{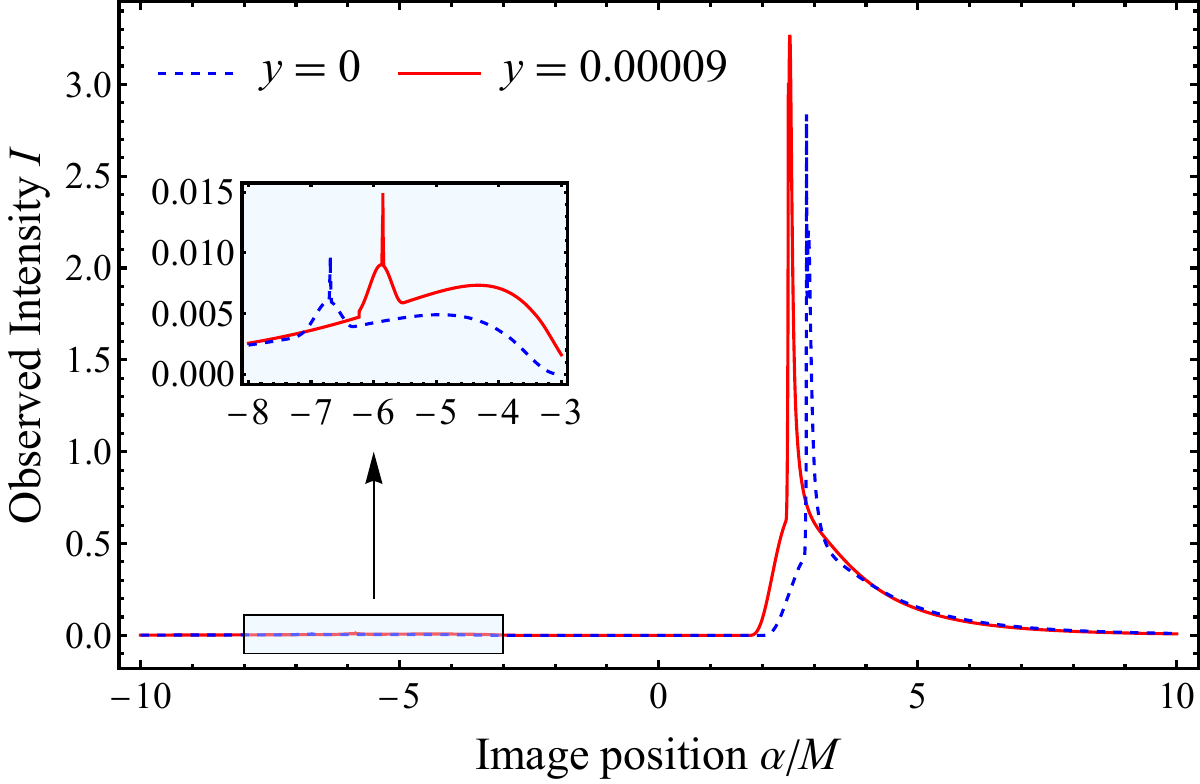}
    \includegraphics[width=0.495\textwidth]{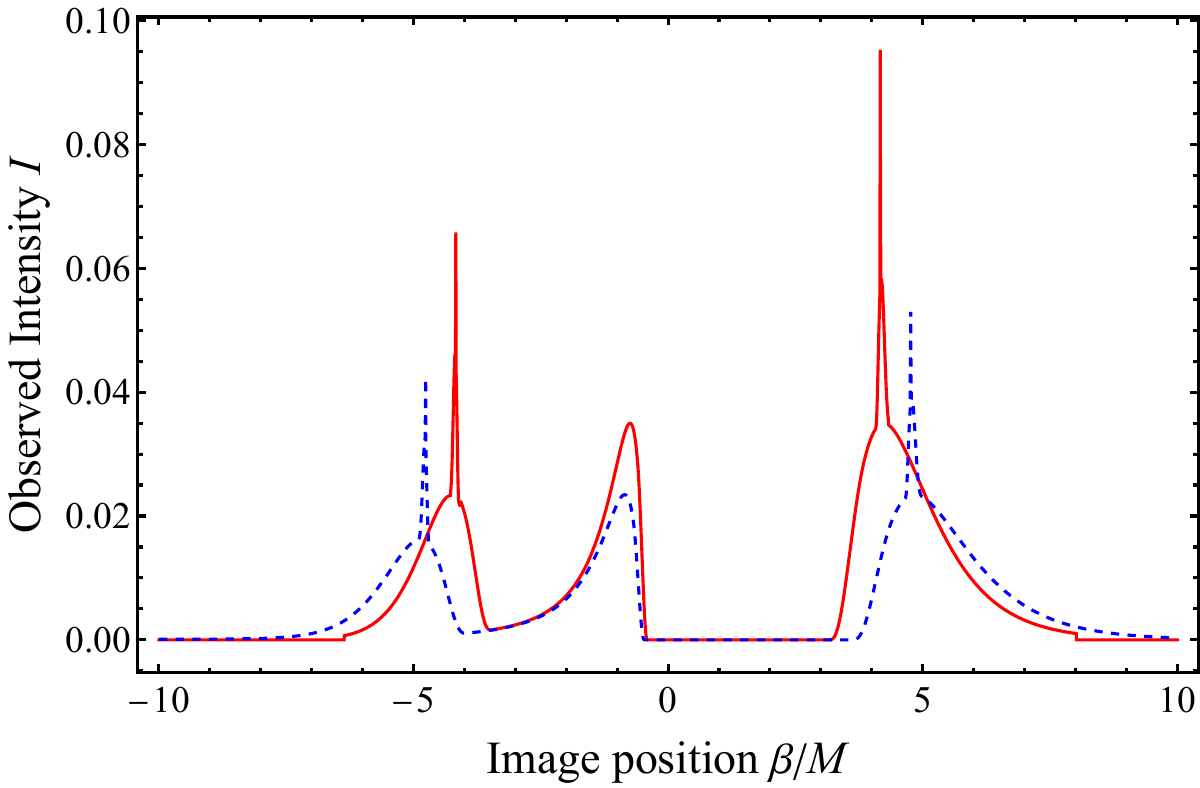}
    \caption{Comparison of the observed intensity with and without a cosmological constant. The black hole spin is fixed at \(a=0.9\)
    The image on the left shows the observed intensity on the \(\alpha\)-axis at \(\beta\) equals 0 by a ZAMO at position \((r_0=50M,\theta_0=75^\circ)\), 
    while the image on the right shows the observed intensity on the \(\beta\)-axis at \(\alpha\) equals 0. The red solid lines are the case of \(y=0.00009\)
    and the blue dashed lines are that of Kerr cases.}
    \label{fig:intensityVy}
\end{figure}
\begin{figure}[htbp]
    \centering
    \includegraphics[width=0.495\textwidth]{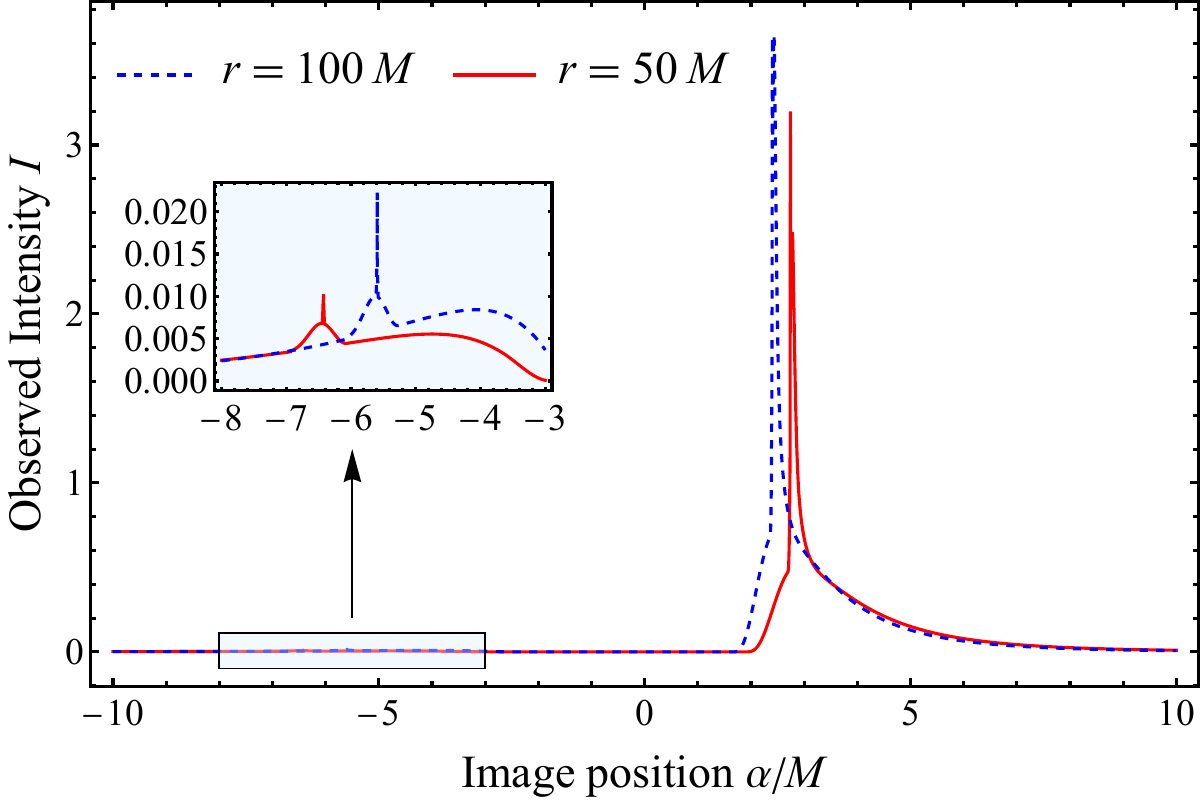}
    \includegraphics[width=0.495\textwidth]{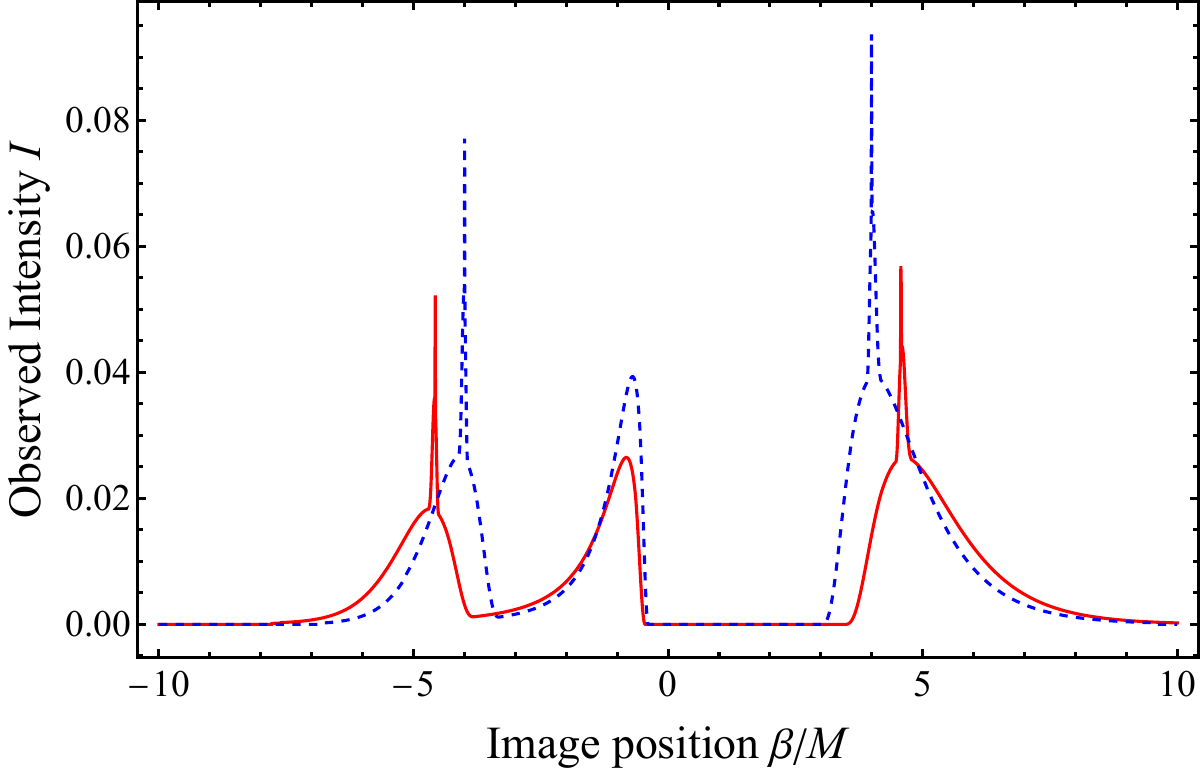}

    \caption{Comparison of the observed intensity at different observation distances. The dimensionless cosmological constant is \(y=0.00003\), 
    and the black hole spin is \(a=0.9\). The corresponding radius of the OSCO is approximately \(180M\).
    The image on the left shows the intensity on the \(\alpha\)-axis at \(\beta=0\) observed by a ZAMO at inclination \((\theta_0=75^\circ)\), 
    while the image on the right shows the observed intensity on the \(\beta\)-axis at \(\alpha=0\).}
    \label{fig:intensityVr}
\end{figure}


Another noteworthy observation from \autoref{fig:bigtosmall} is the increased brightness of the black hole image as the cosmological constant rises. Recalling that \autoref{eq:Itotal} used to calculate the observed intensity does not take into account the cost of radiative transmission and the flux of light received by the observer. Thus, the observed intensity here can be understood as the luminosity of the accretion disk.

In \autoref{fig:intensityVy}, we illustrate brightness cross-sections perpendicular and parallel to the projected spin axis, comparing cases with zero and non-zero cosmological constants. The black hole parameters and observer's position are \(a=0.9M\) and \((r_0=50M, \theta_0=75^\circ)\), respectively. 
It can be seen that the presence of the cosmological constant increases the peak value of the observation intensity, and the positions of the peaks on the left and right sides are closer together. This is consistent with the reduction and brightening of the photon ring shown by \autoref{fig:bigtosmall}.


To gain further insight into this luminance enhancement, we fix the cosmological constant and compare the observed intensity at different observation distances. \autoref{fig:intensityVr} illustrates that an observer at a distance of \(r_0=100M\) will observe a smaller and brighter photon ring than an observer at \(r_0=50M\). It should be noted that the increase in observed intensity shown in \autoref{fig:intensityVy} and \autoref{fig:intensityVr} is always accompanied by the position of the peaks on both sides being close to each other.

This property proposes a theoretically feasible method to test the cosmological constant, which is to image a black hole at different observation distances and analyze the change in the luminosity of the black hole with the change in the observer's position. With current black hole imaging technology, it is not practical to image black holes from beyond Earth. One possible solution is to take advantage of changes in observation distances caused by the relative motion of galaxies. Formulating a practical method is beyond the current scope of this paper.

\section{Conclusion}
\label{6}
In this paper, we have developed a comprehensive analytic method for simulating images of Kerr-de Sitter black holes illuminated by equatorial thin accretion disks. The model of the equatorial accretion disk is established in \autoref{2.3}, assuming particles strictly follow equatorial geodesic motion. The range of stable circular orbits in the equatorial plane has been depicted in \autoref{fig:co-counter-rotating}.

By considering a zero angular momentum observer within the domain of outer communication, we have traced the associated light rays backward through the emitting region to calculate the intensity at each screen position \((\alpha, \beta)\). The formula \autoref{i=fjg} we use to generate the black hole image are fully explicit, as detailed in \autoref{3.2} and \autoref{4}.

Using these explicit expressions, we have simulated images of Kerr-de Sitter black holes illuminated by both prograde and retrograde accretion disks (\autoref{fig:grid-prog} and \autoref{fig:grid-retrog}). We have also investigated the effects of cosmological constants on characteristic curves (\autoref{fig:curves}), relative size (\autoref{fig:bigtosmall}, \autoref{fig:four-shadow} and \autoref{fig:newsize}), and observed intensity (\autoref{fig:intensityVy} and \autoref{fig:intensityVr}).

\begin{figure}[htbp]
    \centering
    \includegraphics[width=\linewidth]{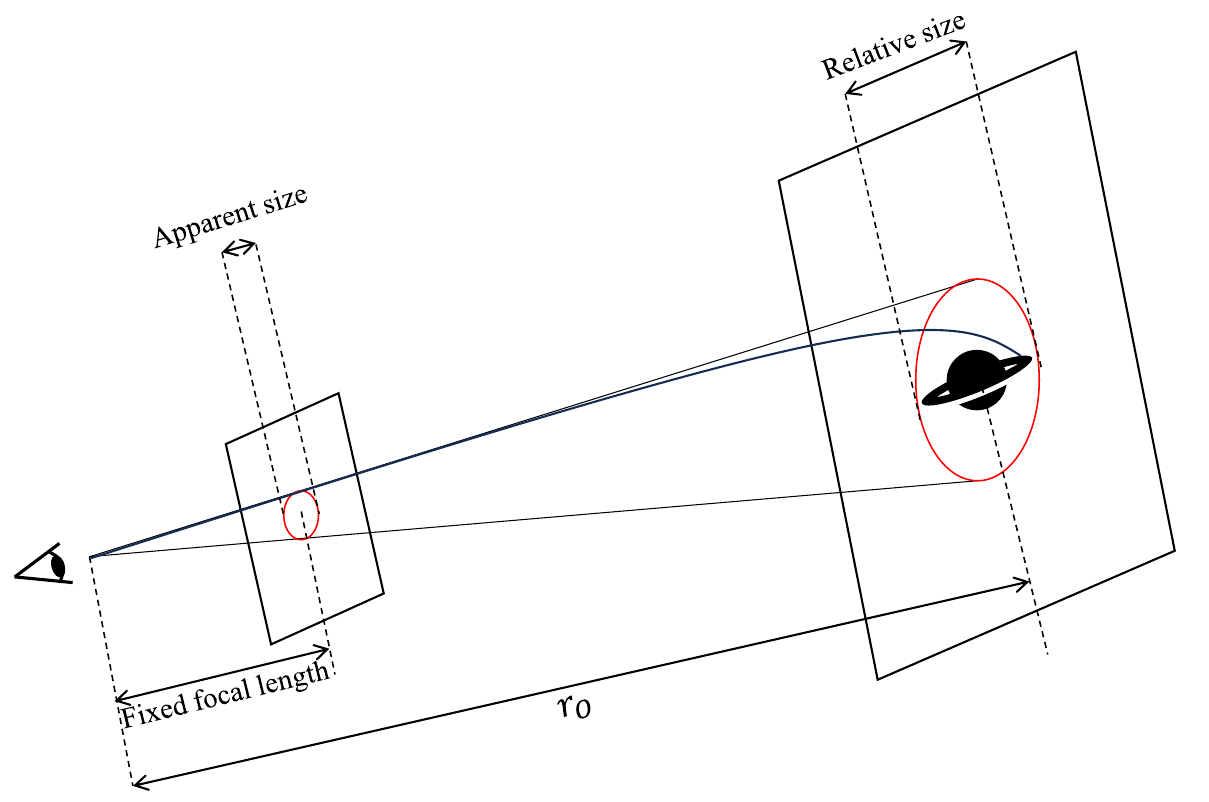}
    \caption{A schematic diagram of the black hole imaging process. The apparent size of a black hole image corresponds to a fixed focal length. The relative size of a black hole image corresponds to a focal length equal to the distance from the observer to the black hole.}
    \label{fig:apparentandrelative}
\end{figure}

Before summarizing our results, we clarify the difference between the relative size and the apparent size of the black hole used in this paper. The apparent size is equivalent to the apparent diameter in astronomy, which is in line with our visual intuition that the farther away an object is, the smaller it looks. When our imaging focal length is a fixed value (for example, the focal length of the human eye is the distance from the lens to the retina), the size of the object in the image can be regarded as the apparent size. In most studies of black hole images, researchers usually set the distance between the observer and the black hole to be infinite. At this time, a fixed focal length will lead to an infinitesimal apparent size. To obtain a black hole image with a finite size, the common practice is to set the focal length as the distance from the observer to the black hole to eliminate the influence of infinite distance on the image size. In this paper, the definition of \autoref{abtoxy} is to set the focal length as the distance from the observer to the black hole, and consequently the size of the black hole image defined by \autoref{eq:rho} is called the relative size. The relative size is equivalent to the actual size of the object in flat spacetime. In curved spacetime it can be understood as the optically measurable size of an object. The distinction between relative size and apparent size is displayed in \autoref{fig:apparentandrelative}.

Our results show that the effects of cosmological constants on black hole images are mainly reflected in two aspects: relative size and observed intensity.
In flat spacetime, changes in observation distances do not affect the relative sizes of objects. In a Kerr spacetime, a distant observer will get similar results as in flat spacetime since the Kerr spacetime is asymptotically flat. However, in Kerr-de Sitter spacetimes with a cosmological constant \(\Lambda >0\), we find that the relative size of the black hole decreases with increasing observational distance and tends to zero as the observer approaches the cosmological horizon \(r_C\) (\autoref{fig:shadowsize}). Furthermore, we find that the observer's position in the universe \(r_0/r_C\) determines the relative size of the black hole (\autoref{fig:newsize}).
Another notable effect of the cosmological constant on black hole images is the increase in luminosity, accompanied by a decrease in relative size (\autoref{fig:intensityVy} and \autoref{fig:intensityVr}).


\acknowledgments
This work is supported by National Science Foundation of China grant No.~11105091 and No.~12175099. We greatly appreciate the comments from Prof. Yongge Ma and Prof. Miao Li for very useful discussions.

\appendix

\section{Solutions of $\lambda(\alpha,\beta)$ and $\eta(\alpha,\beta)$}
\label{appendix:solution}
Given a screen position \((\alpha,\beta)\), the corresponding parameters \(\lambda,\eta\) of null geodesics is determined by \autoref{eqxy}.
Substituting \autoref{RRandTheta} into \autoref{eqxy}, we have
\begin{align}
    \label{eqa:beta}
    \beta(\lambda ,\eta )&=\pm _yr\frac{\sqrt{g_{r r}}\Delta _r}{\sqrt{g_{\theta  \theta }}\Delta _{\theta }}\sqrt{\frac{-\Xi ^2 (\lambda  \csc\theta
        -a \sin\theta)^2+\left(\eta +(a-\lambda )^2\Xi ^2\right) \Delta _{\theta }}{\Xi ^2\left(r^2+a^2-a \lambda \right)^2-\Delta _r\left(\eta
    +\Xi ^2(\lambda -a)^2\right)}} \,,
    \\
    \label{eqa:alpha}
        \alpha(\lambda ,\eta )&=r\frac{\sqrt{g_{r r}}}{\sqrt{g_{\phi  \phi }}}\Delta _r \frac{\lambda }{\sqrt{\Xi ^2\left(r^2+a^2-a \lambda \right)^2-\Delta _r\left(\eta
    +\Xi ^2(\lambda -a)^2\right)}} \,.
\end{align}
Combining \autoref{eqa:alpha} and \autoref{eqa:beta}, we get a quadratic equation of \(\lambda\):
\begin{equation}
    \label{eqa:quadratic}
    A \lambda ^2+B \lambda +C=0 \,,
\end{equation}
where
\begin{align}
        A&=\Xi ^2 r^2 \left( \csc ^2\theta\frac{ \Delta _r}{\Delta _{\theta }}  -  a^2 \Xi ^2 
        +\frac{\left(\beta ^2+r^2\right)\Sigma ^2 \csc ^4\theta   \Delta _r}{\alpha ^2 \left(\left(a^2+r^2\right)^2 \Delta _{\theta }^2 \csc ^2\theta-a^2 \Delta _r\right)}
        \right)  \,,
    \\B&=2 a r^2 \Xi ^2\left(\Xi ^2 \left(a^2+r^2\right)-\frac{\Delta _r}{\Delta _{\theta }}\right)  \,,
    \\C&=\Xi ^2 r^2 \left(a^2 \sin ^2\theta\frac{ \Delta _r}{\Delta _{\theta }}-\Xi ^2 \left(a^2+r^2\right)^2\right) \,.
\end{align}
Solving \autoref{eqa:quadratic} and \autoref{eqa:beta}, and then discarding the irrational solution, we finally obtain
\begin{align}
    \label{eqa:vvv1}
    \lambda(\alpha,\beta) &=-\beth - \sqrt{\beth ^2-
    \frac{a^2 \sin ^2\theta \Delta _r-\Xi ^2 \left(a^2+r^2\right)^2 \Delta _{\theta }}
    {\frac{\left(\beta ^2+r^2\right)\Sigma ^2 \Delta _{\theta } \csc ^4\theta   \Delta _r}
    {\alpha ^2 \left[\left(a^2+r^2\right)^2 \Delta _{\theta }^2 \csc ^2\theta -a^2 \Delta _r\right] }
    -a^2 \Xi ^2 \Delta _{\theta } + \csc ^2\theta  \Delta _r}
    }  \,,
    \\ 
    \label{eqa:vvv2}
    \eta(\alpha,\beta) &=-\frac{\lambda ^2 \Xi ^2 r^2 \Sigma ^2 \csc ^4\theta}
    {\alpha ^2 \left[  \left(a^2+r^2\right)^2 \Delta _{\theta }^2 \csc ^2\theta-a^2 \Delta _r\right]  }
    +\frac{\Xi ^4 \left(a^2-a \lambda +r^2\right)^2}{\Delta _r}-\Xi ^4 (a-\lambda )^2  \,,
\end{align}
where
\begin{align}
    \beth =    \frac{a \left(\Xi ^2 \left(a^2+r^2\right) \Delta _{\theta }-\Delta _r\right)}
    {\frac{\left(\beta ^2+r^2\right)\Sigma ^2 \Delta _{\theta } \csc ^4\theta   \Delta _r}
    {\alpha ^2 \left[\left(a^2+r^2\right)^2 \Delta _{\theta }^2 \csc ^2\theta - a^2 \Delta _r \right] }
    -a^2 \Xi ^2 \Delta _{\theta } + \csc ^2\theta  \Delta _r}
 \,.
\end{align}
In the limit \(\Lambda \to 0\) of Kerr black holes,
\autoref{eqa:vvv1} and \autoref{eqa:vvv2} reduce to the null geodesic parameters related to the ZAMO's screen position in Kerr spacetimes.
In the limit of \(\Lambda \to 0\) and \(r \to \infty\), we return to the case of considering distant observers in Kerr spacetimes. 
The \autoref{eqa:vvv1} and \autoref{eqa:vvv2} reduce to
\begin{align}
    \lambda(\alpha,\beta) &= - \alpha \sin\theta \,,
    \\
    \eta(\alpha,\beta) &=\left(\alpha ^2-a^2\right) \cos \theta +\beta ^2 \,,
\end{align}
which coincide with Eq.(58) and Eq.(59) of \cite{gralla_lensing_2020} (see also \cite{gralla_observational_2018}).

\section{Redshift factor}
\label{appendix:redshift}
In the previous section (\autoref{2.3}), we presented the equatorial accretion model and provided the four-velocity equation for particles at each equatorial radius. In this appendix, we derive the redshift factor as defined in \autoref{eq:redshiftfactor}. 

For \(r_{\text{ISCO}}<r<r_{\text{OSCO}}\), particles within the accretion disk follow circular orbits, and their corresponding four-velocity is given by \autoref{eq:circularU}. The associated redshift factor is calculated as follows:
\begin{equation}
    \label{eqa:gin}
\begin{aligned}
        g_\text{circular} (\lambda,\eta,r) &=\frac{\nu _{\text{obs}}}{\nu _{\text{em}}}=\frac{k_{\mu }u_{\text{obs}}^{\mu }}{k_{\mu }u_{\text{em}}^{\mu }}=\frac{-\zeta +\gamma
         \lambda }{-u^t+\lambda  u^\phi}
    \\&= \frac{(\gamma  \lambda -\zeta ) \pm \sqrt{\left(-ya^2\mp 2 a \sqrt{\frac{1}{r^3}-y}\right)-\frac{3}{r}+1}}
    {\Xi  \left(\left(a \sqrt{\frac{1}{r^3}-y}\pm \lambda  \sqrt{\frac{1}{r^3}-y}\right)\mp 1\right)} \,,
\end{aligned}
\end{equation}
where \(\zeta\) and \(\gamma\) are defined in \autoref{tetrad}:
\begin{align}
    \zeta &=\left. 2 \sqrt{\frac{\Xi ^2 \Sigma  \left(\left(a^2+r^2\right)^2 \Delta _{\theta }^2-a^2 \sin ^2\theta \Delta _r\right)}
    {\Delta _{\theta } \Delta _r \left(a^2 \cos (2 \theta )+a^2+2 r^2\right)^2}} \right|_{(r_0,\theta_0)}  \,,
    \\\gamma &=\left.  \frac{\zeta  \left(a \left(\left(a^2+r^2\right) \Delta _{\theta }-\Delta _r\right)\right)}
    {\left(a^2+r^2\right)^2 \Delta _{\theta }^2-a^2 \sin ^2\theta \Delta _r} \right|_{(r_0,\theta_0)} \,.
\end{align}

For \(r_H<r<r_{\text{ISCO}}\), the redshift factor is
\begin{equation}
    \label{eqa:gout}
    g_\text{infall}(r,\lambda ,\eta,\nu_r)=\frac{-\zeta +\gamma  \lambda }{-u_s^r+\nu _r\frac{\sqrt{R(r)}}{\Delta _r}u_s^r+\lambda  u_s^\phi} \,,\qquad
    \nu _r=\text{Sign}\left[P_r\right] \,,
\end{equation}
where
\begin{align}
    u^t_s &=-g^{00} E_{\text{ISCO}}+g^{03} L_{\text{ISCO}} \,,\qquad          u_s^\phi =-g^{03} E_{\text{ISCO}}+g^{33} L_{\text{ISCO}}   \,,
    \\ u_s^r &=- \sqrt{g^{11}\left(-E_{\text{ISCO}}^2g^{00}+2E_{\text{ISCO}} L_{\text{ISCO}} g^{03} - L_{\text{ISC0}}^2g^{33} - 1\right)}  \,.
\end{align}

\section{null bound geodesic condition}
\label{appendix:critical-curve}

Solving the circular orbit conditions \(\dot{r}=\ddot{r}=0\) in Kerr-de sitter spacetimes, we find that
\begin{equation}
    \begin{aligned}
         \lambda (R) &=\left.\frac{-4r \Delta _r}{a \Delta _r'}+\frac{r^2+a^2}{a}\right|_{r=R} \,,
        \\ \eta  (R)&=\left.\frac{16r^2 \Xi ^2 \Delta _r}{\left(\Delta _r'\right){}^2}-\Xi
        ^2(\lambda -a)^2\right| _{r=R}  \,,
    \end{aligned}
    \label{eq:lambdaAndeta}
\end{equation}
where \(R\) is the radii of bound orbits. The range of \(R\) is determined by an inequality \(\Theta (\theta )\geq 0\) and a implied condition \(r_+<R<r_C\).
Using \autoref{eq:lambdaAndeta}, the condition \(\Theta (\theta )\geq 0\) becomes
\begin{equation}
    \begin{aligned}
        16r^2a^2\sin ^2\theta  \Delta _{\theta } \Delta _r\geq  \left.\left(4r \Delta _r-\Sigma \Delta _r'\right){}^2\right| _{r=R} \,,
    \end{aligned}
\end{equation}
or
\begin{equation}
    \\R\in [r^\gamma_{+},r^\gamma_-] \,,
\end{equation}
where \(r^\gamma_{\pm}\) is the roots of \(\eta=0\) in the range \(r_+<r<r_C\):
\begin{align}
    r^\gamma_{\pm}&=k[1+\sqrt{1-4a^2\Lambda}\cos(\frac{1}{3}\arccos (\frac{\mp|b|}{9M^2} ) ) ] \,, \\
    k&=\frac{6M}{3+4a^2\Lambda} \,, \qquad  
    b=\frac{9M^2-32a^6\Lambda^2-48a^4\Lambda+18a^2(6\Lambda M^2-1)}
    {9M^2(4a^2\Lambda-1)\sqrt{1-4a^2\Lambda}}   \,.
\end{align}

\bibliographystyle{unsrt}
\bibliography{blackholeimage}

\end{document}